\documentclass{article}
\usepackage[preprint]{neurips_2026}

\usepackage[utf8]{inputenc}
\usepackage[T1]{fontenc}
\usepackage{hyperref}
\usepackage{url}
\usepackage{booktabs}      
\usepackage{amsmath,amssymb}
\usepackage{nicefrac}
\usepackage{microtype}
\usepackage{xcolor}
\usepackage{graphicx}
\usepackage{tikz-cd}
\usepackage{graphicx}
\usepackage{float}
\usepackage{float}
\hypersetup{colorlinks=true,allcolors=blue}
\graphicspath{{figures/}}

\newcommand{\nScoredTasks}{228}
\newcommand{\nTestTasks}{272}
\newcommand{\ejbversion}{v0.1}

\title{Bookkeeping Is Not Judgement: EJBench, an Engineering Judgement Benchmark Built from a Conjugate Heat Transfer Design Campaign}

\author{%
  James Sadler\thanks{Corresponding author.} \\
  Gradient Dynamics, London, UK \\
  \texttt{james.sadler@gradient-dynamics.com} \\
  \And
  Rotimi Alabi \\
  Gradient Dynamics, London, UK \\
  \texttt{rotimi.alabi@gradient-dynamics.com} \\
  \And
  Kotub Uddin \\
  AESC, London, UK \\
  \texttt{kotub.uddin@envision-energy.com} \\
}

\begin{document}
\maketitle

\begin{abstract}
We introduce EJBench, a benchmark of engineering judgement: whether a model can predict how a design behaves at new conditions from a few earlier results, and decide on it. Unlike engineering benchmarks that test whether an agent can operate simulation or CAD software, EJBench is software-agnostic and asks what an engineer would conclude without running a simulation. Its tasks come from a conjugate heat transfer design campaign for battery cold plates, in three families: matching unlabelled test results to their operating points, choosing which stored runs to reveal before signing off a design, and ranking new duties by peak cell temperature. Every answer is taken from a converged simulation, not from an LLM judge. The best of ten frontier models, GPT-6 Astra, scores 0.71, and models state almost the same confidence on tasks they get wrong as on tasks they get right. We also post-train Qwen3.5-9B with LoRA and reinforcement learning, rewarded by the share of answer criteria the simulation results confirm; on the held-out test split its score rises from 0.071 to 0.106. We release the train, validation and test splits, the evaluation harness and the leaderboard.
\end{abstract}

\section{Introduction}     
In recent years, LLMs and agentic coding platforms have demonstrated remarkable improvements in software engineering capabilities. The best score on SWE-Bench Pro rose from 23.3\% at release \citep{deng2025swebenchpro} to 64.3\% seven months later \citep{openai2026gpt55}. Coding is unusual in multiple ways that make it perfect for model training: training data is public and abundant, it is easy to measure correctness through tests, and those tests provide exact and quantifiable feedback. These criteria are not met for physical engineering work, where data is often siloed in companies, and physics-verified feedback is limited and expensive to produce. Additionally, engineers must balance numerous requirements and real-world constraints.

Engineers must execute judgement over the design and operation of physical systems to achieve objectives. This discernment is often compounded from years of experience from similar tasks. Most engineering benchmarks measure something else. Many draw their problems from textbooks and examinations \citep{zhou2025engibench,siddique2025physicseval,geissler2025utqa,zheng2026tpscalcbench}, and on thermodynamics calculations frontier models already score above 90\% \citep{duzkar2026thermoqa}. Others test whether a model can operate a simulation tool, and show that runnable is not the same as correct. On PLE, the best agent satisfied 91\% of individual constraints, yet only 41.3\% of its deliverables passed every hard safety and feasibility check \citep{han2026ple}.

In this paper, we introduce EJBench~\ejbversion{}\footnote{\url{https://huggingface.co/datasets/GradientDynamics/EJBench}, git tag \texttt{\ejbversion}.}, whose first suite covers the thermal design of battery cold plates, a new benchmark designed to test agents judgement over engineering design and simulation tasks, built from a real design campaign. EJBench therefore targets judgement, like EngiBench, but scores it against physics, like EngDesign, with exact answers rather than an LLM judge. We keep every converged, quality-checked case, including unusual and poor designs.

\section{Related Work}
\label{sec:related}
 
\subsection{Agentic Coding Benchmarks}
\label{sec:related-coding}
 
HumanEval \citep{chen2021codex} introduced 164 hand-written Python problems checked by unit tests, of which Codex, a GPT-3 model fine-tuned on public GitHub code, solved 28.8\%. LiveCodeBench \citep{jain2025livecodebench} later limited contamination by collecting new contest problems continuously. SWE-bench \citep{jimenez2024swebench} moved evaluation to agents resolving 2,294 real GitHub issues, graded by each repository's own tests, with a human-validated subset in SWE-bench Verified \citep{chowdhury2024swebenchverified}; SWE-Bench Pro \citep{deng2025swebenchpro} and Terminal-Bench \citep{merrill2026terminalbench} extend this to longer and more realistic tasks. All share one property: answers are checked by running them, so grading is exact, cheap and automatic. EJBench brings this property to engineering design, with solved physics simulations in place of unit tests.
 
\subsection{Reinforcement Learning for Agentic Coding}
\label{sec:related-rl}
 
The same checks made coding a productive domain for reinforcement learning. In reinforcement learning with verifiable rewards (RLVR) \citep{lambert2024tulu3}, a programmatic check of the final answer replaces a learned reward model; combined with GRPO \citep{shao2024deepseekmath}, it underlies DeepSeek-R1 \citep{guo2025deepseekr1} and reasoning models that now compete with elite human programmers \citep{elkishky2025competitive}. For agents, the bottleneck is environments: SWE-Gym \citep{pan2025swegym}, R2E-Gym \citep{jain2025r2egym} and SWE-smith \citep{yang2025swesmith} supply thousands of executable tasks with automatic checks for training software agents. Engineering design has no equivalent, because its data sit inside companies and checking a design means running a simulation. Every EJBench answer comes from a solved case, so its training split can serve directly as a reinforcement-learning environment with physics-verified rewards (Section~\ref{sec:experiments}).
 
\subsection{Benchmarks for Agentic Engineering}
\label{sec:related-engineering}
 
Most engineering benchmarks for language models test knowledge and calculation, with problems drawn from textbooks and examinations \citep{siddique2025physicseval,geissler2025utqa,duzkar2026thermoqa,zheng2026tpscalcbench}. EngiBench \citep{zhou2025engibench} adds open-ended problems scored by an LLM against rubrics, and P-1 AI proposed a framework that adapts Bloom's taxonomy to the design of physical systems \citep{neema2025engagi}. Other benchmarks test whether models can operate engineering software such as COMSOL \citep{mudur2025feabench} or OpenFOAM \citep{somasekharan2026cfdllmbench}, or generate CAD \citep{alrashedy2025cadprompt}, and EngDesign \citep{guo2025engdesign} grades 101 design tasks by simulation.
Most recently, RWE-Bench \citep{xplore2026rwebench}, EngiWorld \citep{gao2026engiworld} and PLE \citep{han2026ple} test whole workflows, from CAD and CAE tools to code-compliant deliverables graded by automated checks and LLM rubrics. EJBench complements these: they test whether an agent can carry out an engineering process, while EJBench tests whether a model knows the right answer, which is the judgement engineers use to set up, check and approve that work. Its tasks come from a real design campaign rather than textbooks. Every answer is derived from a converged physics solution, not an LLM judge, and is clear by more than the simulation's numerical uncertainty.

\section{EJBench}
\label{sec:ejbench}

The ground-truth data were produced with Gradient Dynamics' own design and simulation software, but the benchmark does not depend on it. A task gives the model the design and operating data an engineer would have, in plain units, and asks for a decision. It never asks for software syntax, so any model or harness can be scored on it, and the answers could be reproduced with any validated conjugate heat transfer (CHT) solver. This sets it apart from work in which a language model drives a CFD code~\citep{chen2024metaopenfoam,yue2026foamagent,somasekharan2026cfdllmbench}, and from engineering design benchmarks that check an answer by running a simulation ~\citep{guo2025engdesign}: EJBench asks what an engineer would conclude without
running one.

\subsection{Source Data and Splits}
\label{sec:data}

\paragraph{Designs.} EJBench~\ejbversion{} is derived from a simulation campaign of thousands of battery module and cold plate designs at multiple operating conditions. The plates fall into six layouts: centre-fed parallel channels, headered serpentines and headerless serpentines, each with the flow along the plate's length or across its width. Packs hold 6 to 52 cells of one of six cell types: three prismatic LFP cells (EVE LF105, LF280K and A41 class) and three cylindrical cells (18650, 21700 and 4680 format). A seventh cell, the Panasonic CGA463450 prismatic cell, appears in 46 designs only and is held out to test transfer to an unseen cell family, in the spirit of benchmarks built around distribution shift~\citep{koh2021wilds}. The space between the cells is either air, with natural convection and radiation, or potting compound.

\paragraph{Operating points.} The design of experiments (DOE) plans up to 8 operating points for each of the unique geometries. Each design has one chilled high-load point, one thermal-stress point and one hydraulic-exploration point. At its other points the flow is the one a pump controller would set: a thermal controller's choice, held up by the pump's minimum speed or held down by the hydraulic limit. The points use three coolants (water, and 30 and 50~wt\% ethylene glycol), inlet temperatures of 5--35\,\textdegree C, ambient temperatures of 10--45\,\textdegree C, flow rates of 0.025--18\,L/min, and charge or discharge at 0.2C--5.0C, with state of charge 10--95\% and state of health 70--100\%. Cell heat generation is computed from the operating point and applied uniformly; it ranges from 0.08\,W to 3.4\,kW per pack. Most points are normal operating conditions; the three tail points are deliberately extreme.

\paragraph{Simulation and answers.} Every point is a steady CHT simulation in FluxCore, our GPU solver, coupling the coolant, the plate, the cells, the pads, the interstitial air or potting and the surroundings. The coolant is laminar below a Reynolds number of 2,300 in each channel branch and uses the $k$--$\omega$ SST model above it~\citep{menter1994two}. Air is buoyant and laminar, with radiation, and the outer surfaces exchange heat with the ambient through a stated heat-transfer coefficient. A run becomes an answer only if it converges and passes the mass, flow-rate and energy-balance checks. Each answer run yields four quantities used by the tasks: peak cell temperature rise above the coolant inlet (median 8.6\,K, 5th--95th percentile 1.3--22.8\,K), cell temperature spread, coolant temperature rise and plate pressure drop (median 3.0\,kPa).

\paragraph{Splits.} Splits are by geometry: a design and all its operating points sit in one split, and no design or run is shared between splits. The DOE assigns every geometry to a source pool before any simulation is run. Table~\ref{tab:splits} gives the splits. \texttt{test} is the benchmark. It holds \nTestTasks{} tasks, of which a fixed \nScoredTasks{} give the benchmark score
(Section~\ref{sec:scoring}). Its designs were never published before this release, and it includes every Panasonic design with enough runs, tagged \texttt{source\_split = ood\_panasonic}; at 7 tasks they are too few to report as a separate score. \texttt{train} and \texttt{validation} are for fine-tuning. They come from designs in no evaluation split and use the easier training formats of Appendix~\ref{app:training}, so a \texttt{validation} score tracks training, not benchmark performance. A further set of tasks, built in the same way as \texttt{test}, is kept private to detect contamination (\texttt{private\_eval}, Section~\ref{sec:leakage}).

\begin{table}[t]
  \centering
  \caption{EJBench~\ejbversion{} splits, by geometry. \texttt{test} holds \nTestTasks{} tasks; the
  benchmark score uses a fixed subset of \nScoredTasks{} of them (Section~\ref{sec:scoring}), and the
  other 44, all campaign tasks, are released but not scored. \texttt{train} and \texttt{validation}
  will be rebuilt as more simulations finish; \texttt{test} is frozen for this version.}
  \label{tab:splits}
  \begin{tabular}{llrr}
    \toprule
    Split & Use & Tasks & Scored \\
    \midrule
    \texttt{train}      & fine-tuning                   & 1,281         & --              \\
    \texttt{validation} & fine-tuning, held-out designs & 133           & --              \\
    \texttt{test}       & benchmark                     & \nTestTasks{} & \nScoredTasks{} \\
    \bottomrule
  \end{tabular}
\end{table}

\subsection{Task Definition}
\label{sec:tasks}

EJBench measures one skill an engineer uses daily: \emph{predicting a design's behaviour at new conditions from a few earlier results}, and acting on that prediction. Each task shows a design and some completed runs of it, and asks for a decision about operating points the model has not seen. There are three families (Table~\ref{tab:families}).

\begin{table}[t]
  \centering
    \caption{Task families. Chance is the score of uniform random answers. Test is every task in   \texttt{test}; Scored is the subset used for the benchmark score (Section~\ref{sec:scoring}).
  Counts are for EJBench~\ejbversion{}.}
  \label{tab:families}
  \small
  \begin{tabular}{p{2.6cm}p{4.2cm}p{3.6cm}rrr}
    \toprule
    Family & Model sees & Model gives & Chance & Test & Scored \\
    \midrule
    Campaign reconstruction & Five planned test points; six unlabelled result sheets, one misfiled from another plate and some with illegible entries; two held-out duties with temperature limits & Which sheet belongs to each point, which is misfiled, and pass or fail at each duty & 0.0003 & 144 & 100 \\
    Evidence sign-off & Inputs of four or more stored runs, results hidden; three duties with  temperature and pressure-drop limits; a budget of three reveals & Temperature and pressure
      pass or fail at each duty & 0.016 & 80 & 80 \\
    Operating envelope & Three completed runs with results; four new duties & Duties ranked by peak cell rise, and the heat-balance regime of each & 0.0002 & 48 & 48 \\
    \bottomrule
  \end{tabular}
\end{table}

\paragraph{Campaign reconstruction} tests whether a model can tell results that fit a design's physics from results that do not. Sheets must be matched using the relations a design obeys: pressure drop scales with the coolant's friction and flow, the coolant rise follows the heat balance, and the peak rise follows the load, the flow and the ambient offset. The misfiled sheet looks plausible on its own: it comes from another design at a similar pressure drop and peak rise.

\paragraph{Evidence sign-off} tests whether a model knows which evidence it needs. It is the only  multi-step family: the model reveals up to three stored runs with a \texttt{get\_run\_result} tool and then submits. The runs closest to the duties are not the most informative; the runs that span the inputs pin down the response best.

\paragraph{Operating envelope} tests whether a model can extrapolate across conditions. The regimes are defined by the share of the pack's heat that comes from the surroundings: the pack's own heat dominates (\texttt{own\_heat}), a significant ambient gain (\texttt{mixed}), a gain larger than the pack's own heat (\texttt{ambient\_gain}), or a net loss to the surroundings (\texttt{ambient\_loss}). Ranking uses the rise above each duty's own inlet, because absolute temperatures would often be settled by the inlet temperature alone.

\paragraph{Construction rules.} Every number in a prompt is either a DOE input or a stored FluxCore result, rounded as a test sheet would report it (results to three significant figures). Tasks are drawn at random from one design's runs and kept only if they pass three kinds of gate:
\begin{enumerate}
  \item \emph{Solvable.} A reference method that uses only the prompt gets every part right. For temperatures it is a three-term least-squares fit of the peak rise to the heat over the coolant capacity rate, the heat, and the ambient-to-inlet offset; for pressure drop it is a friction-scaling relation; for regimes it is the coolant heat balance. For campaign tasks the best assignment of sheets must also be unique. Every evidence task was checked at generation: a fixed reference method that chooses three runs from the inputs alone, before seeing any result, must get all six verdicts right, or the task is discarded. The limits sit at least one tolerance from the simulated values, and at most half of the possible three-run choices work, so the task is solvable but the choice of runs matters.
  \item \emph{Not answerable by a shortcut.} A quick method must get the task wrong: the closest run, a two-term fit to the two closest runs, or ranking by heat, heat over flow or ambient offset alone (Section~\ref{sec:shortcuts}).
  \item \emph{Balanced.} Pass and fail outcomes and stray-sheet states are drawn in turn, and a
    case is a duty in at most one task per family within a split.
\end{enumerate}

\subsection{Scoring}
\label{sec:scoring}
Every task is scored 1 or 0: 1 only if every part of the answer is right, which for a campaign task means all five links, the misfiled sheet and both verdicts. Each prompt ends with its scoring rule, so the model knows the format and that partial answers earn nothing. A reply that cannot be parsed scores 0. Each part is also recorded, so failures can be traced (Section~\ref{sec:failures}).

The benchmark score is computed on a fixed subset of \nScoredTasks{} of the \nTestTasks{} \texttt{test} tasks: every evidence and envelope task and 100 of the 144 campaign tasks. The other 44 campaign tasks are released with the split but never enter the score. The headline is the macro mean over the three families, so the larger campaign family does not dominate, and its precision is set by the two smaller families; more campaign tasks would add cost but little precision. The 100 are chosen from task metadata only, never from model results, in proportion to the sign-off outcomes and stray-sheet states and spread over as many designs as possible. The list and its SHA-256 hash ship with the release, and a score over all \nTestTasks{} tasks is not the benchmark score. We report a 95\% percentile bootstrap interval~\citep{efron1979bootstrap} from 1,000 resamples of the scored tasks, as recommended for model evaluations~\citep{miller2024adding}.

\subsection{An Example Task}
\label{sec:example}

Figure~\ref{fig:example} abridges a public \texttt{train} task. The pack is six 105\,Ah LFP cells in air on a headerless serpentine plate. Ranking the duties by heat generation gives Y, X, W, Z, and that ranking is wrong: duty~Z has the least heat but the warmest coolant and surroundings, and its ambient offset of 10.4\,K adds more heat than the pack generates. Its true rise, 8.2\,K, puts it above duty~W (3.6\,K). The three completed runs fix the three-term fit exactly, and the fit gives the right order, Y, X, Z, W. The coolant heat balances of the runs show that ambient heat entered at every run, which gives the regimes: X and Z \texttt{ambient\_gain}, W and Y \texttt{mixed}.

\begin{figure}[t]
  \centering
  \fbox{\begin{minipage}{0.95\linewidth}\footnotesize
  You are reviewing the operating envelope of a battery cold plate from three completed test runs.
  You may not run a simulation. Rises are above that run's coolant inlet.\\[2pt]
  \textbf{Design.} Pack: 6 $\times$ EVE LF105-class prismatic cell (105\,Ah, LFP/graphite) in 6 rows
  $\times$ 1 column; a 1.2\,mm thermal-interface pad under each cell; open air between cells.
  Cold plate: aluminium headerless serpentine, 157 $\times$ 268 $\times$ 10.5\,mm, 3 turns, channel
  diameter 2.37\,mm. \ldots\\[2pt]
  \textbf{Completed runs.} 1: \ldots \quad 2: \ldots \quad 3: water at 18.1\,\textdegree C,
  0.200\,L/min; ambient 22.2\,\textdegree C; 3.58\,W. \emph{Result: peak cell rise 3.27\,K; cell
  spread 2.48\,K; coolant rise 0.736\,K; pressure drop 7.86\,kPa.}\\[2pt]
  \textbf{New duties.} W: 30\,wt\% glycol at 10.5\,\textdegree C, 0.173\,L/min; ambient
  14.2\,\textdegree C; 7.59\,W. \quad X: 30\,wt\% glycol at 21.7\,\textdegree C, 0.200\,L/min;
  ambient 36.9\,\textdegree C; 14.6\,W. \quad Y: 30\,wt\% glycol at 14.4\,\textdegree C,
  0.192\,L/min; ambient 28.0\,\textdegree C; 34.9\,W. \quad Z: water at 34.1\,\textdegree C,
  0.200\,L/min; ambient 44.5\,\textdegree C; 6.62\,W.\\[2pt]
  \textbf{Heat-balance regimes.} own\_heat, mixed, ambient\_gain, ambient\_loss (definitions given).\\[2pt]
  Rank the four duties by peak cell rise above their own coolant inlet, largest first, and give
  the regime of each duty. \emph{Scoring: the score is 1 only if the ranking and every regime are
  correct, and 0 otherwise.}\\[2pt]
  \textbf{Answer.} Ranking Y, X, Z, W (rises 14.4, 12.6, 8.2, 3.6\,K); regimes X and Z
  ambient\_gain, W and Y mixed.
  \end{minipage}}
  \caption{An abridged \texttt{operating\_envelope} task from the public \texttt{train} split
  (\texttt{ejb-training-0.1-train-envelope-02009-bc219e}). The full prompt gives every coolant
  property, the load and the external heat-transfer coefficient of each run and duty.}
  \label{fig:example}
\end{figure}

\section{Making the Benchmark Valid}
\label{sec:validity}

\subsection{Answer validity}
\label{sec:answer-validity}

\begin{figure}[t]
  \centering
  \includegraphics[width=\linewidth]{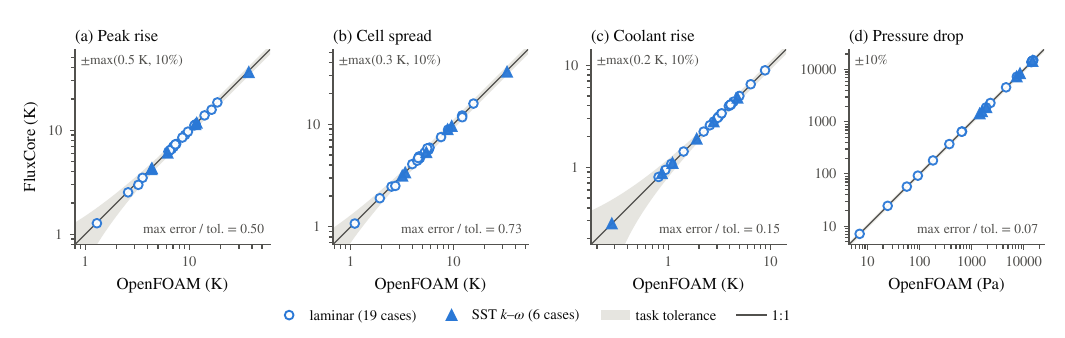}
  \caption{FluxCore against OpenFOAM for the four task quantities: rises in kelvin, pressure drop as relative error. The grey bands mark the task tolerance.}
  \label{fig:parity}
\end{figure}

Each answer is a simulation result, so the benchmark is only as good as the solver. Two checks apply.
\begin{enumerate}
  \item \emph{Run-level checks.} Every answer run converged and passed mass, flow-rate and energy balance checks (Section~\ref{sec:data}); runs that converged but failed them are excluded, not repaired.
    \item \emph{Independent solver.} We solve 25 verification cases in OpenFOAM
    v2512~\citep{weller1998tensorial} with \texttt{chtMultiRegionSimpleFoam} on the same mesh as the stored answer run (2.5--9.5 million cells), with the same boundary conditions, material properties, turbulence model and wall treatment. Cases 1--19 are laminar (branch Reynolds number 24--579) and cases 20--25 turbulent (2,330--3,470); 23 are potted and one has air between the cells. We compare the four task quantities as errors on rises and relative errors on pressure drop, because absolute temperatures near 300\,K hide large errors in a rise of a few kelvin, and divide each error by its task tolerance. No error exceeds its tolerance (Figure~\ref{fig:parity}). The largest, 0.92 of the tolerance, is a pressure drop in a  turbulent case; laminar pressure drops agree to within 0.5\%, coolant rises to within 0.07\,K, and FluxCore's peak rise is 1.6\% below OpenFOAM's on average. Appendix~\ref{app:verification} provides more detail on verification.
  
\end{enumerate}

\subsection{Shortcut audits}
\label{sec:shortcuts}

Benchmarks can often be solved by cues unrelated to the skill they claim to measure~\citep{gururangan2018annotation,poliak2018hypothesis,mccoy2019right,geirhos2020shortcut}. EJBench guards against this in two ways.

\paragraph{By construction.} Each family names the shortcuts a model could take, and a task is kept only if they fail (Section~\ref{sec:tasks}). In \texttt{operating\_envelope}, the true order is given neither by ranking on heat, heat over flow or ambient offset alone, nor by the closest run, a one-term fit, or a fit that ignores the ambient offset. Every campaign task has at least one duty that the full fit gets right and a quick method gets wrong, and at least one ambient-heated duty. In \texttt{evidence\_signoff}, revealing the runs closest to the duties gives a wrong temperature verdict, and at most half of all possible choices of three runs give every verdict. The reference method, which uses only the prompt, scores 1 on every task.

\paragraph{By measurement.} The best constant answer, chosen with knowledge of the test answers, scores at most 0.03 on campaign tasks (no assignment of sheets repeats more than twice in 100 campaign tasks), 0.0625 on evidence tasks and 0.042 on envelope tasks. Uniform random answers score 0.0003, 0.016 and 0.0002.

\subsection{Leakage and contamination}
\label{sec:leakage}

\paragraph{Leakage between splits.} Splits are by geometry. The DOE assigned every geometry to a source pool before any simulation ran, and the v0.1 allocation of designs to splits is published with the data. No design appears in two splits. The misfiled sheet in a campaign task is drawn from another design in the same split, so a test task never shows a training run.

\paragraph{Contamination of models.} The test designs and their results were never public before this release, and the prompts are generated from stored results rather than collected from the web, so a model can only have seen a test answer if the release itself was in its training data. Contamination of this kind is common and hard to detect after the fact~\citep{sainz2023nlp,golchin2024time,xu2024benchmark}. Every released record carries a canary string with a unique identifier, following BIG-Bench~\citep{srivastava2023beyond}, so that it can be filtered from training corpora and its presence in a model can be tested; \citet{jacovi2023stop} argue for such safeguards on any released test data. To detect contamination that does happen, we keep \texttt{private\_eval}: 58 tasks built by the same generator from designs that are never released. A model whose \texttt{test} score is well above its \texttt{private\_eval} score has probably seen the test set~\citep{oren2024proving}. Because tasks are drawn from stored results, new draws can replace a compromised test set without new simulations, much as refreshed benchmarks do~\citep{kiela2021dynabench,white2025livebench}.

\section{Experiments}
\label{sec:experiments}
\subsection{Agent Harness}
\label{sec:harness}

Every model runs the released harness, \texttt{eval/run\_model.py}. A single-turn task sends the prompt with a JSON output schema and parses the last JSON object in the reply. An \texttt{evidence\_signoff} task runs a turn loop: each reply is one JSON tool call, \texttt{get\_run\_result} returns a stored run until the budget of three is spent, and \texttt{submit} ends the episode, with at most eight turns. API models answer each task once at high reasoning effort; Qwen3.5-9B answers four times, with a reply limit of 128,000 tokens including reasoning (131,072 for DeepSeek V4.1 Flash and Muse Spark 1.3). Qwen3.5-9B runs on our own server with thinking on, temperature 0.6, top-$p$ 0.95, top-$k$ 20 and a 65,536-token limit. A reply cut off at the limit cannot be parsed and scores 0. Each run records its reasoning effort, whether thinking is on, and the hash of its task set, so scores on different task sets never mix.

\subsection{Zero-shot evaluation}
\label{sec:zeroshot}

\begin{figure}[t]
  \centering
  \includegraphics[width=\linewidth]{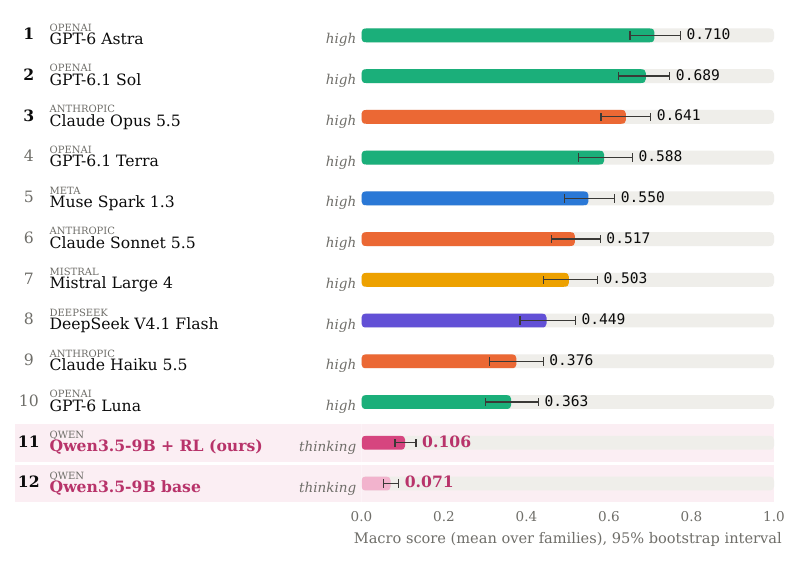}
  \caption{EJBench~\ejbversion{} leaderboard on the \nScoredTasks{} scored test tasks. Bars give the macro score and whiskers its 95\% bootstrap interval; colour marks the developer. API models run once at high reasoning effort. The highlighted rows are Qwen3.5-9B before and after training on EJBench (Section~\ref{sec:training}), with thinking on, each averaged over four runs.}
  \label{fig:leaderboard}
\end{figure}

We evaluate 11 configurations from six developers on the launch leaderboard of November 2026 (Table~\ref{tab:results}, Figure~\ref{fig:leaderboard}). 

\paragraph{No model is close to the ceiling.} GPT-6 Astra leads with a macro score of 0.710. The ten models span 0.36 to 0.71. Qwen3.5-9B, the open model we train in Appendix~\ref{app:training}, scores 0.071. Of the \nScoredTasks{} tasks, 24 are solved by none of the ten models (14 campaign, 9 evidence, 1 envelope) and 21 by all of them, although the reference method solves every task from the prompt alone.

\paragraph{Families.} Envelope tasks are the easiest family for every model. The three families rank the models in much the same order (Spearman correlation 0.65 to 0.86), with one clear exception: Claude Sonnet 5.5 is fourth on campaign tasks but last of the ten models on evidence tasks (0.263). Much of that gap is length: in 31 of its 80 evidence tasks a reply reached the 128,000-token limit.

\begin{figure}[t]
  \centering
  \includegraphics[width=\linewidth]{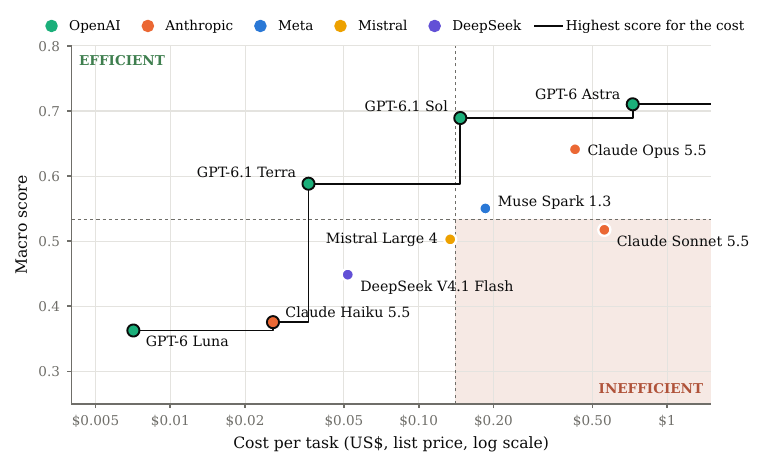}
  \caption{Macro score against cost per task at list price (log scale). The black line joins the
  models with the highest score for the cost. Dashed lines mark the medians; the shaded corner holds
  models that cost more and score less than the median. Qwen3.5-9B is self-hosted and not shown.}
  \label{fig:cost}
\end{figure}

\paragraph{Cost.} Figure~\ref{fig:cost} plots score against cost per task at list price. Cost spans a factor of about 100, from \$0.007 (GPT-6 Luna) to \$0.73 (GPT-6 Astra). Five models set the highest score for the cost: GPT-6 Luna, Claude Haiku 5.5, GPT-6.1 Terra, GPT-6.1 Sol and GPT-6 Astra. GPT-6.1 Terra scores 0.588 at \$0.036 per task, above four models that cost more, and GPT-6.1 Sol comes within 0.02 of GPT-6 Astra at a fifth of the cost. Cost follows the number of tokens written as much as the price per token: Claude Sonnet 5.5 writes twice as many output tokens as Claude Opus 5.5 (9.2M against 4.6M), so it costs more per task (\$0.56 against \$0.43) and scores
0.12 lower.

\paragraph{Calibration.} Brier scores of the models range from 0.196 to 0.361. Qwen3.5-9B states a mean confidence of 0.88 while solving 6\% of tasks, for a Brier score of 0.742.

\begin{table}[t]
  \centering
  \caption{Results on the \nScoredTasks{} scored EJBench~\ejbversion{} test tasks, ranked by macro score. Macro is the mean over families with a 95\% bootstrap interval. Brier is the mean squared difference between stated confidence and score. Cost is per task at list price, including reasoning tokens. API models run at high reasoning effort and answer each task once; the two Qwen3.5-9B~\citep{qwen2026qwen35} rows run with thinking on and average four runs, and the first is the model trained on EJBench (Section~\ref{sec:training}). Qwen3.5-9B is self-hosted, so it has no list price. Model reports: Claude Opus 5.5~\citep{anthropic2026system}, DeepSeek V4.1~\citep{deepseekai2026deepseekv4}.}
  \label{tab:results}
  \small
  \setlength{\tabcolsep}{4pt}
  \begin{tabular}{lcccccr}
    \toprule
    Model & Macro & Campaign & Evidence & Envelope & Brier & \$/task \\
    \midrule
    GPT-6 Astra         & 0.710 {\footnotesize[0.651, 0.773]} & 0.710 & 0.650 & 0.771 & 0.196 & 0.727 \\
    GPT-6.1 Sol         & 0.689 {\footnotesize[0.624, 0.747]} & 0.680 & 0.637 & 0.750 & 0.215 & 0.147 \\
    Claude Opus 5.5     & 0.641 {\footnotesize[0.581, 0.701]} & 0.590 & 0.563 & 0.771 & 0.230 & 0.426 \\
    GPT-6.1 Terra       & 0.588 {\footnotesize[0.527, 0.657]} & 0.490 & 0.588 & 0.688 & 0.216 & 0.036 \\
    Muse Spark 1.3      & 0.550 {\footnotesize[0.492, 0.614]} & 0.430 & 0.450 & 0.771 & 0.296 & 0.185 \\
    Claude Sonnet 5.5   & 0.517 {\footnotesize[0.460, 0.579]} & 0.540 & 0.263 & 0.750 & 0.254 & 0.559 \\
    Mistral Large 4     & 0.503 {\footnotesize[0.440, 0.572]} & 0.350 & 0.450 & 0.708 & 0.352 & 0.134 \\
    DeepSeek V4.1 Flash & 0.449 {\footnotesize[0.384, 0.518]} & 0.350 & 0.413 & 0.583 & 0.361 & 0.052 \\
    Claude Haiku 5.5    & 0.376 {\footnotesize[0.310, 0.441]} & 0.260 & 0.325 & 0.542 & 0.211 & 0.026 \\
    GPT-6 Luna          & 0.363 {\footnotesize[0.300, 0.429]} & 0.180 & 0.388 & 0.521 & 0.350 & 0.007 \\
    \midrule
    Qwen3.5-9B + RL (ours) & 0.106 {\footnotesize[0.081, 0.132]} & 0.018 & 0.097 & 0.203 & 0.712 & -- \\
    Qwen3.5-9B             & 0.071 {\footnotesize[0.054, 0.090]} & 0.022 & 0.075 & 0.115 & 0.742 & -- \\
    \bottomrule
  \end{tabular}
\end{table}

\subsection{Where models fail}
\label{sec:failures}

\paragraph{Which part fails.} Table~\ref{tab:parts} shows how often each part of the answer is fully
right. Weaker models fail campaign tasks mostly at matching sheets: GPT-6 Luna gets every link right
in 28\% of tasks, against 83\% for GPT-6 Astra. In evidence tasks every model does better on the
pressure verdicts than on the temperature verdicts, and Claude Sonnet 5.5's low evidence score comes
from both (36\% and 55\%). In envelope tasks 91 to 97\% of individual regimes are right, but one
wrong regime fails a task about as often as a wrong ranking does.

\begin{table}[t]
  \centering
  \caption{Share of the \nScoredTasks{} scored test tasks in which each part of the answer is fully
  right.}
  \label{tab:parts}
  \small
  \begin{tabular}{lccccccc}
    \toprule
    & \multicolumn{3}{c}{Campaign} & \multicolumn{2}{c}{Evidence} & \multicolumn{2}{c}{Envelope} \\
    \cmidrule(lr){2-4}\cmidrule(lr){5-6}\cmidrule(lr){7-8}
    Model & Links & Misfiled & Verdicts & Temp. & Pressure & Ranking & Regimes \\
    \midrule
    GPT-6 Astra         & 0.83 & 0.87 & 0.83 & 0.75 & 0.86 & 0.90 & 0.88 \\
    GPT-6.1 Sol         & 0.81 & 0.87 & 0.82 & 0.79 & 0.81 & 0.92 & 0.81 \\
    Claude Opus 5.5     & 0.72 & 0.85 & 0.76 & 0.74 & 0.75 & 0.85 & 0.88 \\
    GPT-6.1 Terra       & 0.63 & 0.73 & 0.74 & 0.70 & 0.78 & 0.85 & 0.83 \\
    Muse Spark 1.3      & 0.71 & 0.77 & 0.62 & 0.64 & 0.75 & 0.88 & 0.88 \\
    Claude Sonnet 5.5   & 0.77 & 0.82 & 0.72 & 0.36 & 0.55 & 0.92 & 0.83 \\
    Mistral Large 4     & 0.59 & 0.67 & 0.56 & 0.56 & 0.81 & 0.85 & 0.81 \\
    DeepSeek V4.1 Flash & 0.57 & 0.64 & 0.61 & 0.54 & 0.76 & 0.77 & 0.75 \\
    Claude Haiku 5.5    & 0.50 & 0.61 & 0.48 & 0.40 & 0.78 & 0.73 & 0.75 \\
    GPT-6 Luna          & 0.28 & 0.52 & 0.52 & 0.53 & 0.76 & 0.77 & 0.69 \\
    Qwen3.5-9B          & 0.05 & 0.24 & 0.38 & 0.23 & 0.33 & 0.35 & 0.20 \\
    \bottomrule
  \end{tabular}
\end{table}

\paragraph{The pressure drop hides the misfiled sheet.} A campaign task's misfiled sheet may be complete, or may have lost its coolant rise or its pressure drop, as any sheet can; each case covers about a third of tasks. Over the ten models, the misfiled sheet is found in 85\% of tasks when it has lost its pressure drop, against 67 to 68\% when it keeps it, and the whole task is solved in 62\% against 38\%. Losing the coolant rise makes no difference, so the models are misled by the pressure drop more than they are helped by the heat balance.

\paragraph{Choosing evidence is not the bottleneck.} Over the models, evidence tasks are solved in 52\% of episodes that revealed exactly the reference set, 55\% of episodes that revealed the three runs nearest the duties, and 45\% of the rest. Which runs a model reveals matters little; the models fail at using the evidence, not at choosing it.

\paragraph{Confidence carries little information.} Most models state almost the same confidence on tasks they solve as on tasks they fail. Mistral Large 4 states 0.77 on both, and the largest gap is 0.14 (GPT-6.1 Terra, 0.63 against 0.49). A user cannot tell a right answer from a wrong one by the stated confidence. This matches reports that verbalised confidence is overconfident and that reasoning models are no better calibrated~\citep{xiong2024can,zeng2025thinking}.

\subsection{Training on EJBench}
\label{sec:training}

To test whether the data can improve the engineering judgement of an open model, we post-train Qwen3.5-9B~\citep{qwen2026qwen35} on the released training set and compare it with the untrained model on \texttt{test}.

\paragraph{Training data.} Training draws from the pool of 1,281 \texttt{train} tasks, and checkpoints are compared on the 133 \texttt{validation} tasks (Table~\ref{tab:training-set}). The training formats are single-turn:
campaign tasks with four or five test points, envelope tasks with three or four duties, and \texttt{duty\_signoff}, a sign-off with every shown run's results visible. No training task needs the \texttt{get\_run\_result} tool, so \texttt{evidence\_signoff} is a task layout the model never trains on.
Each task's answer defines a list of criteria: in a campaign task, each test-point link, the misfiled
sheet and each verdict; in a sign-off task, each temperature and pressure verdict; in an envelope task,
the ranking, which must be exact, and each regime. The benchmark scorer checks every criterion against
the simulation results, so no human or model grader is involved. The reward is the fraction of criteria
met; the benchmark itself counts a task as solved only when every criterion is met.

\paragraph{Protocol.} We train a LoRA adapter~\citep{hu2022lora} with reinforcement learning from verifiable rewards~\citep{lambert2024tulu3,guo2025deepseekr1}: group-normalised advantages as in GRPO~\citep{shao2024deepseekmath}, a sequence-level clipped ratio as in GSPO~\citep{zheng2025gspo}, and a small KL term towards the base model. Each step samples 8 replies to each of 8 tasks, with thinking on and replies of up to 32,768 tokens, and the three families get equal shares of the prompts. Before training we fixed the budget at 160 steps (1,280 prompts) on one GPU, and fixed the final checkpoint as the one evaluated on \texttt{test}. Prompts are drawn in a seeded random order, balanced so that each family receives a third of them. The envelope family is the smallest, so its tasks repeat while some campaign and sign-off tasks are never drawn; a full pass over the balanced prompts would take 232 steps. Base and trained models then answer every scored test task four times with the harness settings of Section~\ref{sec:harness}, including the 65,536-token limit. Full settings are in Appendix~\ref{app:training}.

\paragraph{Results.} Training raises the macro full-pass score from 0.071 to 0.106 (change $+0.035$,
95\% interval $+0.006$ to $+0.066$) and the share of criteria met from 0.527 to 0.580 ($+0.030$ to
$+0.076$; Table~\ref{tab:training}). The trained model still ranks below every API model
(Figure~\ref{fig:leaderboard}). Envelope tasks gain most, from 11.5\% to 20.3\% fully right.
Evidence sign-off barely moves (criteria 0.639 to 0.648): likely because its training format is easier, with every run's results shown, about 2.5 criteria a task and none of the traps of the test layout, which has six criteria. Part of the gain is the model learning to finish: 49 of 912 base replies reach the token limit, against 6 of 912 after training, and without those replies the criteria gain is $+0.040$, about three quarters of the full gain.

\begin{table}[t]
  \centering
  \caption{Qwen3.5-9B before and after training, on the \nScoredTasks{} scored test tasks, four runs each. Criteria is the mean share of criteria met per reply; full pass is the share of replies with every criterion met, with replies in parentheses.}
  \label{tab:training}
  \small
  \begin{tabular}{lrcccc}
    \toprule
    & & \multicolumn{2}{c}{Criteria} & \multicolumn{2}{c}{Full pass} \\
    \cmidrule(lr){3-4}\cmidrule(lr){5-6}
    Family & Tasks & Base & Trained & Base & Trained \\
    \midrule
    Campaign & 100 & 0.397 & 0.434 & 2.2\% (9/400)   & 1.8\% (7/400) \\
    Evidence & 80  & 0.639 & 0.648 & 7.5\% (24/320)  & 9.7\% (31/320) \\
    Envelope & 48  & 0.546 & 0.656 & 11.5\% (22/192) & 20.3\% (39/192) \\
    \midrule
    Macro    & 228 & 0.527 & 0.580 & 7.1\% & 10.6\% \\
    \bottomrule
  \end{tabular}
\end{table}

\paragraph{Campaign tasks barely move.} Training raised the campaign criteria score from 0.397 to 0.434, but not the full-pass rate: 9 of 400 base replies (2.2\%) were fully right, against 7 of 400 (1.8\%) after training. Four of the base model's nine come from one task, which it passes in all four runs and the trained model in none; on the other 99 tasks the trained model passes 7 replies against 5. In that task the misfiled sheet has the largest peak rise and the heaviest-load point is not on it, so only the coolant heat balance rules it out. Across the 54 test tasks with this trap, both models fall for it equally often (35\% base, 34\% trained), so training did not teach a shortcut. The limit is finding the misfiled sheet: the base model finds it in 24.5\% of replies and the trained model in 27.0\%, against 17\% by chance, and \texttt{validation} shows no change (0.30 and 0.30). Training helped only when the misfiled sheet had lost its pressure drop (0.31 to 0.41), the case the API models also find easiest (Section~\ref{sec:failures}); it did not help when the sheet had lost its coolant rise (0.20 to 0.20) or was complete (0.23 to 0.19). Below a full pass, answers did improve: replies with five or more of the eight criteria wrong fell from 240 to 205 of 400, and link accuracy rose from 0.344 to 0.398.

The reward explains why. GRPO learns from differences between replies to the same prompt, and 82\% of campaign training groups held no fully right reply. On training prompts the campaign criteria score rose to about 0.46 by step 40 and stayed there to step 160, with 2--5\% of replies fully right. The misfiled sheet is one of eight criteria and a full pass earns no bonus, so finding it is worth an eighth of the reward. Family balancing also gave campaign tasks only a third of the prompts: 434 of the 617 campaign tasks were seen once each and 183 (30\%) never.

\section{Release and Maintenance}
\label{sec:release}

We release version~\ejbversion{} on Hugging Face
(\url{https://huggingface.co/datasets/GradientDynamics/EJBench}): the \texttt{train}, \texttt{validation} and \texttt{test} splits, the harness and scorer (\texttt{eval/}), the leaderboard task set and its hash, and every model run reported here, with a datasheet
(Appendix~\ref{app:datasheet};~\citealp{gebru2021datasheets}) and Croissant
metadata~\citep{akhtar2024croissant}. The release is licensed under CC BY-NC 4.0, which allows reuse for non-commercial purposes with attribution. The generator, the stored simulation fields and \texttt{private\_eval} are withheld. Questions and leaderboard entries go to the corresponding author.

\paragraph{Versions.} Each version is a tagged release. Within a version, \texttt{test} and its hash do not change, so leaderboard scores stay comparable. \texttt{train} and \texttt{validation} will be rebuilt as more simulations finish; a rebuild never changes \texttt{test}. A change to \texttt{test}, such as removing tasks whose margins prove smaller than the measured solver uncertainty, makes a new version, and the leaderboard records which version each score is on.

\paragraph{Leaderboard.} An entry must answer every scored task with the released harness and report its settings. Only complete runs are ranked, and each entry lists its date, cost and token use.

\section{Limitations}
\label{sec:limitations}

\begin{itemize}
  \item \textbf{Simulation, not experiment.} Answers come from one solver, checked against a second solver but not against measurements. Cell heat is applied uniformly from a heat model.
  \item \textbf{All-or-nothing scores.} They make tasks hard to guess but hide near misses; the recorded parts show where an answer went wrong.
  \item \textbf{Templated prompts.} The numbers change in every task, but the layout does not, and a model could learn it from the released splits.
  \item \textbf{One sample per task.} Intervals cover the choice of tasks, not run-to-run variation, and reasoning effort means different things for different providers.
\end{itemize}

\section{Next Steps}
\label{sec:next}

\paragraph{Data.} \texttt{train} and \texttt{validation} will grow as more simulations finish, and the provisional tolerances will be replaced. A training sign-off format that matches \texttt{evidence\_signoff}, with hidden runs and six criteria, may move the family that training left unchanged. Because tasks are drawn from stored results, a new \texttt{test} can be drawn without new simulations if the current one is compromised; it would be released as a new
version.

\paragraph{Training policy.} The campaign results point to the reward. A bonus for a full pass would separate a fully right answer from one that misses only the misfiled sheet, and logging each part of the answer during training would show which parts move. A warm start on fully right campaign replies, chosen by rejection sampling, would give GRPO groups that contain full passes, and oversampling training tasks with the heat-balance trap would target the main failure. A full pass over the pool and
a second seed would firm up the data-scaling result. Every change uses training data only, with checkpoints chosen on \texttt{validation}.

\section{Conclusion}
\label{sec:conclusion}

We introduce EJBench, a benchmark of engineering judgement: whether a model can predict how a design behaves at new conditions from a few earlier results, and decide on it. Its tasks are built on battery cold plates, and every answer comes from a conjugate heat transfer simulation verified against OpenFOAM. The best model, GPT-6 Astra, scores 0.71, and 24 of the \nScoredTasks{} scored tasks are solved by none of the ten API models, although every task can be solved from the prompt alone. Models state almost the same confidence on tasks they get wrong as on tasks they get right, so their answers cannot yet be trusted without a check. Price is a poor guide to score: GPT-6.1 Terra beats four dearer models at \$0.04 per task. Training an open 9B model on the released data raises its score from 0.071 to 0.106, with the largest gain on envelope tasks, but barely moves campaign tasks, where finding the misfiled sheet remains the limit. We release the data, harness and runs so that progress on this kind of judgement can be measured.

\bibliographystyle{plainnat}
\bibliography{references}

@article{chen2021codex,
  title   = {Evaluating Large Language Models Trained on Code},
  author  = {Chen, Mark and Tworek, Jerry and Jun, Heewoo and Yuan, Qiming and Pinto, Henrique Ponde de Oliveira and Kaplan, Jared and Edwards, Harri and Burda, Yuri and Joseph, Nicholas and Brockman, Greg and others},
  journal = {arXiv preprint arXiv:2107.03374},
  year    = {2021}
}

@inproceedings{jain2025livecodebench,
  title     = {{LiveCodeBench}: Holistic and Contamination Free Evaluation of Large Language Models for Code},
  author    = {Jain, Naman and Han, King and Gu, Alex and Li, Wen-Ding and Yan, Fanjia and Zhang, Tianjun and Wang, Sida and Solar-Lezama, Armando and Sen, Koushik and Stoica, Ion},
  booktitle = {The Thirteenth International Conference on Learning Representations (ICLR)},
  year      = {2025}
}

@inproceedings{jimenez2024swebench,
  title     = {{SWE}-bench: Can Language Models Resolve Real-World {GitHub} Issues?},
  author    = {Jimenez, Carlos E. and Yang, John and Wettig, Alexander and Yao, Shunyu and Pei, Kexin and Press, Ofir and Narasimhan, Karthik},
  booktitle = {The Twelfth International Conference on Learning Representations (ICLR)},
  year      = {2024}
}

@misc{chowdhury2024swebenchverified,
  title        = {Introducing {SWE-bench Verified}},
  author       = {Chowdhury, Neil and Aung, James and {Chan Jun Shern} and Jaffe, Oliver and Sherburn, Dane and Starace, Giulio and Mays, Evan and Dias, Rachel and Aljubeh, Marwan and Glaese, Mia and others},
  howpublished = {\url{https://openai.com/index/introducing-swe-bench-verified/}},
  year         = {2024},
  note         = {OpenAI, 13 August 2024. Accessed 2026-10-19}
}

@article{deng2025swebenchpro,
  title   = {{SWE-Bench Pro}: Can {AI} Agents Solve Long-Horizon Software Engineering Tasks?},
  author  = {Deng, Xiang and Da, Jeff and Pan, Edwin and He, Yannis Yiming and Ide, Charles and Garg, Kanak and Lauffer, Niklas and Park, Andrew and Pasari, Nitin and Rane, Chetan and others},
  journal = {arXiv preprint arXiv:2509.16941},
  year    = {2025}
}

@inproceedings{merrill2026terminalbench,
  title     = {{Terminal-Bench}: Benchmarking Agents on Hard, Realistic Tasks in Command Line Interfaces},
  author    = {Merrill, Mike A. and Shaw, Alexander G. and Carlini, Nicholas and Li, Boxuan and Raj, Harsh and Bercovich, Ivan and Shi, Lin and Shin, Jeong Yeon and Walshe, Thomas and Buchanan, E. Kelly and others},
  booktitle = {The Fourteenth International Conference on Learning Representations (ICLR)},
  year      = {2026}
}

@misc{openai2026gpt55,
  title        = {Introducing {GPT-5.5}},
  author       = {{OpenAI}},
  howpublished = {\url{https://openai.com/index/introducing-gpt-5-5/}},
  year         = {2026},
  note         = {23 April 2026. Accessed 2026-10-19}
}

@inproceedings{lambert2024tulu3,
  author        = {Lambert, Nathan and Morrison, Jacob and Pyatkin, Valentina and Huang, Shengyi and Ivison, Hamish and Brahman, Faeze and Miranda, Lester James V. and Liu, Alisa and Dziri, Nouha and Lyu, Shane and Gu, Yuling and Malik, Saumya and Graf, Victoria and Hwang, Jena D. and Yang, Jiangjiang and Le Bras, Ronan and Tafjord, Oyvind and Wilhelm, Chris and Soldaini, Luca and Smith, Noah A. and Wang, Yizhong and Dasigi, Pradeep and Hajishirzi, Hannaneh},
  title         = {{T{\"u}lu 3}: Pushing Frontiers in Open Language Model Post-Training},
  booktitle     = {Second Conference on Language Modeling (COLM)},
  year          = {2025},
  url           = {https://arxiv.org/abs/2411.15124},
}

@article{shao2024deepseekmath,
  title   = {{DeepSeekMath}: Pushing the Limits of Mathematical Reasoning in Open Language Models},
  author  = {Shao, Zhihong and Wang, Peiyi and Zhu, Qihao and Xu, Runxin and Song, Junxiao and Bi, Xiao and Zhang, Haowei and Zhang, Mingchuan and Li, Y. K. and Wu, Y. and others},
  journal = {arXiv preprint arXiv:2402.03300},
  year    = {2024}
}

@article{guo2025deepseekr1,
  title   = {{DeepSeek-R1} Incentivizes Reasoning in {LLMs} through Reinforcement Learning},
  author  = {Guo, Daya and Yang, Dejian and Zhang, Haowei and Song, Junxiao and Wang, Peiyi and others},
  journal = {Nature},
  volume  = {645},
  number  = {8081},
  pages   = {633--638},
  year    = {2025},
  doi     = {10.1038/s41586-025-09422-z}
}

@article{elkishky2025competitive,
  title   = {Competitive Programming with Large Reasoning Models},
  author  = {El-Kishky, Ahmed and Wei, Alexander and Saraiva, Andre and Minaiev, Borys and Selsam, Daniel and Dohan, David and Song, Francis and Lightman, Hunter and Clavera, Ignasi and others},
  journal = {arXiv preprint arXiv:2502.06807},
  year    = {2025}
}

@inproceedings{pan2025swegym,
  title     = {Training Software Engineering Agents and Verifiers with {SWE-Gym}},
  author    = {Pan, Jiayi and Wang, Xingyao and Neubig, Graham and Jaitly, Navdeep and Ji, Heng and Suhr, Alane and Zhang, Yizhe},
  booktitle = {Proceedings of the 42nd International Conference on Machine Learning},
  series    = {Proceedings of Machine Learning Research},
  volume    = {267},
  pages     = {47717--47737},
  year      = {2025}
}

@article{jain2025r2egym,
  title   = {{R2E-Gym}: Procedural Environments and Hybrid Verifiers for Scaling Open-Weights {SWE} Agents},
  author  = {Jain, Naman and Singh, Jaskirat and Shetty, Manish and Zheng, Liang and Sen, Koushik and Stoica, Ion},
  journal = {arXiv preprint arXiv:2504.07164},
  year    = {2025}
}

@inproceedings{yang2025swesmith,
  title     = {{SWE-smith}: Scaling Data for Software Engineering Agents},
  author    = {Yang, John and Lieret, Kilian and Jimenez, Carlos E. and Wettig, Alexander and Khandpur, Kabir and Zhang, Yanzhe and Hui, Binyuan and Press, Ofir and Schmidt, Ludwig and Yang, Diyi},
  booktitle = {Advances in Neural Information Processing Systems (Datasets and Benchmarks Track)},
  volume    = {38},
  year      = {2025}
}

@article{zhou2025engibench,
  title   = {{EngiBench}: A Benchmark for Evaluating Large Language Models on Engineering Problem Solving},
  author  = {Zhou, Xiyuan and Wang, Xinlei and He, Yirui and Wu, Yang and Zou, Ruixi and Cheng, Yuheng and Xie, Yulu and Liu, Wenxuan and Zhao, Huan and Xu, Yan and Gu, Jinjin and Zhao, Junhua},
  journal = {arXiv preprint arXiv:2509.17677},
  year    = {2025}
}

@inproceedings{siddique2025physicseval,
  title     = {{PhysicsEval}: Inference-Time Techniques to Improve the Reasoning Proficiency of Large Language Models on Physics Problems},
  author    = {Siddique, Oshayer and Alam, J. M Areeb Uzair and Rafy, Md Jobayer Rahman and Raiyan, Syed Rifat and Mahmud, Hasan and Hasan, Md Kamrul},
  booktitle = {Findings of the Association for Computational Linguistics: IJCNLP-AACL 2025},
  year      = {2025}
}

@article{geissler2025utqa,
  title   = {From Canonical to Complex: Benchmarking {LLM} Capabilities in Undergraduate Thermodynamics},
  author  = {Gei{\ss}ler, Anna and Bien, Luca-Sophie and Sch{\"o}ppler, Friedrich and Hertel, Tobias},
  journal = {arXiv preprint arXiv:2508.21452},
  year    = {2025}
}

@article{duzkar2026thermoqa,
  title   = {{ThermoQA}: A Three-Tier Benchmark for Evaluating Thermodynamic Reasoning in Large Language Models},
  author  = {D{\"u}zkar, Kemal},
  journal = {arXiv preprint arXiv:2604.19758},
  year    = {2026}
}

@article{zheng2026tpscalcbench,
  title   = {{TPS-CalcBench}: A Benchmark and Diagnostic Evaluation Framework for {LLM} Analytical Calculation Competence in Hypersonic Thermal Protection System Engineering},
  author  = {Zheng, Jinglai and Qiao, Chuhan and Huang, Haiming},
  journal = {arXiv preprint arXiv:2604.17966},
  year    = {2026}
}

@article{neema2025engagi,
  title   = {On the Evaluation of Engineering Artificial General Intelligence},
  author  = {Neema, Sandeep and Jha, Susmit and Nagel, Adam and Lew, Ethan and Sureshkumar, Chandrasekar and Gordic, Aleksa and Shimmin, Chase and Nguygen, Hieu and Eremenko, Paul},
  journal = {arXiv preprint arXiv:2505.10653},
  year    = {2025}
}

@article{mudur2025feabench,
  title   = {{FEABench}: Evaluating Language Models on Multiphysics Reasoning Ability},
  author  = {Mudur, Nayantara and Cui, Hao and Venugopalan, Subhashini and Raccuglia, Paul and Brenner, Michael P. and Norgaard, Peter},
  journal = {arXiv preprint arXiv:2504.06260},
  year    = {2025}
}

@article{somasekharan2026cfdllmbench,
  title   = {{CFDLLMBench}: A Benchmark Suite for Evaluating Large Language Models in Computational Fluid Dynamics},
  author  = {Somasekharan, Nithin and Yue, Ling and Cao, Yadi and Li, Weichao and Emami, Patrick and Bhargav, Pochinapeddi Sai and Acharya, Anurag and Xie, Xingyu and Pan, Shaowu},
  journal = {Journal of Data-centric Machine Learning Research},
  volume  = {3},
  number  = {13},
  pages   = {1--40},
  year    = {2026}
}

@inproceedings{alrashedy2025cadprompt,
  title     = {Generating {CAD} Code with Vision-Language Models for {3D} Designs},
  author    = {Alrashedy, Kamel and Tambwekar, Pradyumna and Zaidi, Zulfiqar Haider and Langwasser, Megan and Xu, Wei and Gombolay, Matthew},
  booktitle = {International Conference on Learning Representations (ICLR)},
  year      = {2025}
}

@inproceedings{guo2025engdesign,
  author        = {Guo, Xingang and
               Li, Yaxin and
               Kong, Xiangyi and
               Jiang, Yilan and
               Zhao, Xiayu and
               Gong, Zhihua and
               Zhang, Yufan and
               Li, Daixuan and
               Sang, Tianle and
               Zhu, Beixiao and
               Jun, Gregory and
               Huang, Yingbing and
               Liu, Yiqi and
               Xue, Yuqi and
               Kundu, Rahul Dev and
               Lim, Qi Jian and
               Zhao, Yizhou and
               Granger, Luke Alexander and
               Younis, Mohamed Badr and
               Keivan, Darioush and
               Sabharwal, Nippun and
               Sinha, Shreyanka and
               Agarwal, Prakhar and
               Vandyck, Kojo and
               Mai, Hanlin and
               Wang, Zichen and
               Venkatesh, Aditya and
               Barik, Ayush and
               Yang, Jiankun and
               Yue, Chongying and
               He, Jingjie and
               Wang, Libin and
               Xu, Licheng and
               Chen, Hao and
               Wang, Jinwen and
               Xu, Liujun and
               Shetty, Rushabh and
               Guo, Ziheng and
               Song, Dahui and
               Jha, Manvi and
               Liang, Weijie and
               Yan, Weiman and
               Zhang, Bryan and
               Karnoor, Sahil Bhandary and
               Zhang, Jialiang and
               Pandya, Rutva and
               Gong, Xinyi and
               Ganesh, Mithesh Ballae and
               Shi, Feize and
               Xu, Ruiling and
               Zhang, Yifan and
               Ouyang, Yanfeng and
               Qin, Lianhui and
               Rosenbaum, Elyse and
               Snyder, Corey and
               Seiler, Peter and
               Dullerud, Geir and
               Zhang, Xiaojia Shelly and
               Cheng, Zuofu and
               Hanumolu, Pavan Kumar and
               Huang, Jian and
               Kulkarni, Mayank and
               Namazifar, Mahdi and
               Zhang, Huan and
               Hu, Bin},
  title         = {Toward Engineering {AGI}: Benchmarking the Engineering Design Capabilities of {LLMs}},
  booktitle     = {Advances in Neural Information Processing Systems 38 (NeurIPS 2025), Datasets and Benchmarks Track},
  pages         = {78582--78642},
  year          = {2025},
  doi           = {10.52202/085713-2372},
  eprint        = {2509.16204},
  archivePrefix = {arXiv}
}

@misc{xplore2026rwebench,
  title        = {{RWE-bench}: Real World Engineering Bench},
  author       = {{Xplore}},
  howpublished = {\url{https://xplore.ai/rwe-bench}},
  year         = {2026},
  note         = {Released 27 October 2026. Accessed 2026-11-05}
}

@article{gao2026engiworld,
  title   = {{EngiWorld}: What Can Frontier Agents Deliver in Professional Engineering Environments?},
  author  = {Gao, Hongcheng and Qu, Hailong and Lei, Yu and Sun, Henghui and Li, Haoyang and Wei, Yipeng and Xue, Naihao and Yu, Xiaohan and Tao, Zhuo and Zang, Yihe and others},
  journal = {arXiv preprint arXiv:2609.37686},
  year    = {2026}
}

@article{han2026ple,
  title   = {{PLE}: Physical-Engineering's Last Exam},
  author  = {Han, Rujun and Zhuang, Yufan and Ray, Arijit and Yu, Han and CuiZhu, Zoey and Lee, George and Ming, Yifei and Roy, Binz and Sakalkar, Varun and Tirumalashetty, Vishy and others},
  journal = {arXiv preprint arXiv:2610.30537},
  year    = {2026}
}

@misc{chen2024metaopenfoam,
  author        = {Chen, Yuxuan and
               Zhu, Xu and
               Zhou, Hua and
               Ren, Zhuyin},
  title         = {{MetaOpenFOAM}: an {LLM}-based multi-agent framework for {CFD}},
  year          = {2024},
  howpublished  = {arXiv preprint arXiv:2407.21320},
  eprint        = {2407.21320},
  archivePrefix = {arXiv}
}

@article{yue2026foamagent,
  author        = {Yue, Ling and
               Somasekharan, Nithin and
               Zhang, Tingwen and
               Cao, Yadi and
               Chen, Zhangze and
               Di, Shimin and
               Pan, Shaowu},
  title         = {{Foam-Agent}: A large language model-based multi-agent framework for automating computational fluid dynamics workflows},
  journal       = {Computer Methods in Applied Mechanics and Engineering},
  volume        = {461},
  pages         = {119271},
  year          = {2026},
  doi           = {10.1016/j.cma.2026.119271},
  eprint        = {2505.04997},
  archivePrefix = {arXiv}
}

@inproceedings{koh2021wilds,
  author        = {Koh, Pang Wei and
               Sagawa, Shiori and
               Marklund, Henrik and
               Xie, Sang Michael and
               Zhang, Marvin and
               Balsubramani, Akshay and
               Hu, Weihua and
               Yasunaga, Michihiro and
               Phillips, Richard Lanas and
               Gao, Irena and
               Lee, Tony and
               David, Etienne and
               Stavness, Ian and
               Guo, Wei and
               Earnshaw, Berton and
               Haque, Imran and
               Beery, Sara M. and
               Leskovec, Jure and
               Kundaje, Anshul and
               Pierson, Emma and
               Levine, Sergey and
               Finn, Chelsea and
               Liang, Percy},
  title         = {{WILDS}: A Benchmark of in-the-Wild Distribution Shifts},
  booktitle     = {Proceedings of the 38th International Conference on Machine Learning},
  series        = {Proceedings of Machine Learning Research},
  volume        = {139},
  pages         = {5637--5664},
  publisher     = {PMLR},
  year          = {2021},
  eprint        = {2012.07421},
  archivePrefix = {arXiv}
}

@article{menter1994two,
  author  = {Menter, F. R.},
  title   = {Two-equation eddy-viscosity turbulence models for engineering applications},
  journal = {AIAA Journal},
  volume  = {32},
  number  = {8},
  pages   = {1598--1605},
  year    = {1994},
  doi     = {10.2514/3.12149}
}

@article{weller1998tensorial,
  author  = {Weller, H. G. and Tabor, G. and Jasak, H. and Fureby, C.},
  title   = {A tensorial approach to computational continuum mechanics using object-oriented techniques},
  journal = {Computers in Physics},
  volume  = {12},
  number  = {6},
  pages   = {620--631},
  year    = {1998},
  doi     = {10.1063/1.168744}
}

@article{efron1979bootstrap,
  author        = {Efron, Bradley},
  title         = {Bootstrap Methods: Another Look at the Jackknife},
  journal       = {The Annals of Statistics},
  volume        = {7},
  number        = {1},
  pages         = {1--26},
  year          = {1979},
  doi           = {10.1214/aos/1176344552}
}

@misc{miller2024adding,
  author        = {Miller, Evan},
  title         = {Adding Error Bars to Evals: A Statistical Approach to Language Model Evaluations},
  year          = {2024},
  howpublished  = {arXiv preprint arXiv:2411.00640},
  eprint        = {2411.00640},
  archivePrefix = {arXiv}
}

@inproceedings{gururangan2018annotation,
  author        = {Gururangan, Suchin and
               Swayamdipta, Swabha and
               Levy, Omer and
               Schwartz, Roy and
               Bowman, Samuel R. and
               Smith, Noah A.},
  title         = {Annotation Artifacts in Natural Language Inference Data},
  booktitle     = {Proceedings of the 2018 Conference of the North American Chapter of the Association for Computational Linguistics: Human Language Technologies, Volume 2 (Short Papers)},
  pages         = {107--112},
  publisher     = {Association for Computational Linguistics},
  year          = {2018},
  doi           = {10.18653/v1/N18-2017},
  eprint        = {1803.02324},
  archivePrefix = {arXiv}
}

@inproceedings{poliak2018hypothesis,
  author        = {Poliak, Adam and
               Naradowsky, Jason and
               Haldar, Aparajita and
               Rudinger, Rachel and
               Van Durme, Benjamin},
  title         = {Hypothesis Only Baselines in Natural Language Inference},
  booktitle     = {Proceedings of the Seventh Joint Conference on Lexical and Computational Semantics (*{SEM})},
  pages         = {180--191},
  publisher     = {Association for Computational Linguistics},
  year          = {2018},
  doi           = {10.18653/v1/S18-2023},
  eprint        = {1805.01042},
  archivePrefix = {arXiv}
}

@inproceedings{mccoy2019right,
  author        = {McCoy, R. Thomas and
               Pavlick, Ellie and
               Linzen, Tal},
  title         = {Right for the Wrong Reasons: Diagnosing Syntactic Heuristics in Natural Language Inference},
  booktitle     = {Proceedings of the 57th Annual Meeting of the Association for Computational Linguistics},
  pages         = {3428--3448},
  publisher     = {Association for Computational Linguistics},
  year          = {2019},
  doi           = {10.18653/v1/P19-1334},
  eprint        = {1902.01007},
  archivePrefix = {arXiv}
}

@article{geirhos2020shortcut,
  author        = {Geirhos, Robert and
               Jacobsen, J{\"o}rn-Henrik and
               Michaelis, Claudio and
               Zemel, Richard and
               Brendel, Wieland and
               Bethge, Matthias and
               Wichmann, Felix A.},
  title         = {Shortcut learning in deep neural networks},
  journal       = {Nature Machine Intelligence},
  volume        = {2},
  number        = {11},
  pages         = {665--673},
  year          = {2020},
  doi           = {10.1038/s42256-020-00257-z},
  eprint        = {2004.07780},
  archivePrefix = {arXiv}
}

@inproceedings{sainz2023nlp,
  author        = {Sainz, Oscar and
               Campos, Jon Ander and
               Garc{\'\i}a-Ferrero, Iker and
               Etxaniz, Julen and
               Lopez de Lacalle, Oier and
               Agirre, Eneko},
  title         = {{NLP} Evaluation in trouble: On the Need to Measure {LLM} Data Contamination for each Benchmark},
  booktitle     = {Findings of the Association for Computational Linguistics: {EMNLP} 2023},
  pages         = {10776--10787},
  publisher     = {Association for Computational Linguistics},
  year          = {2023},
  doi           = {10.18653/v1/2023.findings-emnlp.722},
  eprint        = {2310.18018},
  archivePrefix = {arXiv}
}

@inproceedings{golchin2024time,
  author        = {Golchin, Shahriar and
               Surdeanu, Mihai},
  title         = {Time Travel in {LLMs}: Tracing Data Contamination in Large Language Models},
  booktitle     = {The Twelfth International Conference on Learning Representations ({ICLR})},
  pages         = {43008--43029},
  year          = {2024},
  eprint        = {2308.08493},
  archivePrefix = {arXiv}
}

@misc{xu2024benchmark,
  author        = {Xu, Cheng and
               Guan, Shuhao and
               Greene, Derek and
               Kechadi, M-Tahar},
  title         = {Benchmark Data Contamination of Large Language Models: A Survey},
  year          = {2024},
  howpublished  = {arXiv preprint arXiv:2406.04244},
  eprint        = {2406.04244},
  archivePrefix = {arXiv}
}

@article{srivastava2023beyond,
  author        = {Srivastava, Aarohi and
               Rastogi, Abhinav and
               Rao, Abhishek and
               Shoeb, Abu Awal Md and
               Abid, Abubakar and
               Fisch, Adam and
               Brown, Adam R. and
               Santoro, Adam and
               Gupta, Aditya and
               Garriga-Alonso, Adri{\`a} and
               others},
  title         = {Beyond the Imitation Game: Quantifying and extrapolating the capabilities of language models},
  journal       = {Transactions on Machine Learning Research},
  year          = {2023},
  note          = {\url{https://openreview.net/forum?id=uyTL5Bvosj}},
  eprint        = {2206.04615},
  archivePrefix = {arXiv}
}

@inproceedings{jacovi2023stop,
  author        = {Jacovi, Alon and
               Caciularu, Avi and
               Goldman, Omer and
               Goldberg, Yoav},
  title         = {Stop Uploading Test Data in Plain Text: Practical Strategies for Mitigating Data Contamination by Evaluation Benchmarks},
  booktitle     = {Proceedings of the 2023 Conference on Empirical Methods in Natural Language Processing},
  pages         = {5075--5084},
  publisher     = {Association for Computational Linguistics},
  year          = {2023},
  doi           = {10.18653/v1/2023.emnlp-main.308},
  eprint        = {2305.10160},
  archivePrefix = {arXiv}
}

@inproceedings{oren2024proving,
  author        = {Oren, Yonatan and
               Meister, Nicole and
               Chatterji, Niladri and
               Ladhak, Faisal and
               Hashimoto, Tatsunori B.},
  title         = {Proving Test Set Contamination in Black-Box Language Models},
  booktitle     = {The Twelfth International Conference on Learning Representations ({ICLR})},
  pages         = {16354--16372},
  year          = {2024},
  eprint        = {2310.17623},
  archivePrefix = {arXiv}
}

@inproceedings{kiela2021dynabench,
  author        = {Kiela, Douwe and
               Bartolo, Max and
               Nie, Yixin and
               Kaushik, Divyansh and
               Geiger, Atticus and
               Wu, Zhengxuan and
               Vidgen, Bertie and
               Prasad, Grusha and
               Singh, Amanpreet and
               Ringshia, Pratik and
               Ma, Zhiyi and
               Thrush, Tristan and
               Riedel, Sebastian and
               Waseem, Zeerak and
               Stenetorp, Pontus and
               Jia, Robin and
               Bansal, Mohit and
               Potts, Christopher and
               Williams, Adina},
  title         = {{Dynabench}: Rethinking Benchmarking in {NLP}},
  booktitle     = {Proceedings of the 2021 Conference of the North American Chapter of the Association for Computational Linguistics: Human Language Technologies},
  pages         = {4110--4124},
  publisher     = {Association for Computational Linguistics},
  year          = {2021},
  doi           = {10.18653/v1/2021.naacl-main.324},
  eprint        = {2104.14337},
  archivePrefix = {arXiv}
}

@inproceedings{white2025livebench,
  author        = {White, Colin and
               Dooley, Samuel and
               Roberts, Manley and
               Pal, Arka and
               Feuer, Benjamin and
               Jain, Siddhartha and
               Shwartz-Ziv, Ravid and
               Jain, Neel and
               Saifullah, Khalid and
               Dey, Sreemanti and
               {Shubh-Agrawal} and
               Sandha, Sandeep Singh and
               Naidu, Siddartha and
               Hegde, Chinmay and
               LeCun, Yann and
               Goldstein, Tom and
               Neiswanger, Willie and
               Goldblum, Micah},
  title         = {{LiveBench}: A Challenging, Contamination-Limited {LLM} Benchmark},
  booktitle     = {The Thirteenth International Conference on Learning Representations ({ICLR})},
  pages         = {91595--91631},
  year          = {2025},
  eprint        = {2406.19314},
  archivePrefix = {arXiv}
}

@misc{anthropic2026system,
  author        = {{Anthropic}},
  title         = {System Card: {Claude Opus 5.5}},
  howpublished  = {\url{https://www.anthropic.com/claude-opus-5-5-system-card}},
  month         = sep,
  year          = {2026},
}

@article{deepseekai2026deepseekv4,
  author        = {{DeepSeek-AI}},
  title         = {{DeepSeek-V4}: Towards Highly Efficient Million-Token Context Intelligence},
  journal       = {arXiv preprint arXiv:2606.19348},
  year          = {2026},
  eprint        = {2606.19348},
  archivePrefix = {arXiv},
  url           = {https://arxiv.org/abs/2606.19348},
}

@misc{qwen2026qwen35,
  author        = {{Qwen Team}},
  title         = {{Qwen3.5}: Towards Native Multimodal Agents},
  howpublished  = {\url{https://qwen.ai/blog?id=qwen3.5}},
  note          = {Model weights: \url{https://huggingface.co/Qwen/Qwen3.5-9B}},
  month         = feb,
  year          = {2026},
}

@inproceedings{xiong2024can,
  author        = {Xiong, Miao and Hu, Zhiyuan and Lu, Xinyang and Li, Yifei and Fu, Jie and He, Junxian and Hooi, Bryan},
  title         = {Can {LLMs} Express Their Uncertainty? An Empirical Evaluation of Confidence Elicitation in {LLMs}},
  booktitle     = {The Twelfth International Conference on Learning Representations},
  year          = {2024},
  url           = {https://arxiv.org/abs/2306.13063},
}

@inproceedings{zeng2025thinking,
  author        = {Zeng, Qingcheng and Xuan, Weihao and Cui, Leyang and Voigt, Rob},
  title         = {Thinking Out Loud: Do Reasoning Models Know When They're Right?},
  booktitle     = {Proceedings of the 2025 Conference on Empirical Methods in Natural Language Processing},
  pages         = {1394--1407},
  year          = {2025},
  doi           = {10.18653/v1/2025.emnlp-main.73},
}

@inproceedings{hu2022lora,
  author        = {Hu, Edward J. and Shen, Yelong and Wallis, Phillip and Allen-Zhu, Zeyuan and Li, Yuanzhi and Wang, Shean and Wang, Lu and Chen, Weizhu},
  title         = {{LoRA}: Low-Rank Adaptation of Large Language Models},
  booktitle     = {The Tenth International Conference on Learning Representations},
  year          = {2022},
  url           = {https://openreview.net/forum?id=nZeVKeeFYf9},
}

@article{zheng2025gspo,
  author        = {Zheng, Chujie and Liu, Shixuan and Li, Mingze and Chen, Xiong-Hui and Yu, Bowen and Gao, Chang and Dang, Kai and Liu, Yuqiong and Men, Rui and Yang, An and Zhou, Jingren and Lin, Junyang},
  title         = {Group Sequence Policy Optimization},
  journal       = {arXiv preprint arXiv:2507.18071},
  year          = {2025},
}

@misc{vonwerra2020trl,
  author        = {von Werra, Leandro and Belkada, Younes and Tunstall, Lewis and Beeching, Edward and Thrush, Tristan and Lambert, Nathan and Huang, Shengyi and Rasul, Kashif and Gallou{\'e}dec, Quentin},
  title         = {{TRL}: Transformers Reinforcement Learning},
  howpublished  = {\url{https://github.com/huggingface/trl}},
  note          = {GitHub repository},
  year          = {2020},
}

@inproceedings{kwon2023vllm,
  author        = {Kwon, Woosuk and Li, Zhuohan and Zhuang, Siyuan and Sheng, Ying and Zheng, Lianmin and Yu, Cody Hao and Gonzalez, Joseph E. and Zhang, Hao and Stoica, Ion},
  title         = {Efficient Memory Management for Large Language Model Serving with {PagedAttention}},
  booktitle     = {Proceedings of the 29th Symposium on Operating Systems Principles (SOSP)},
  pages         = {611--626},
  year          = {2023},
  doi           = {10.1145/3600006.3613165},
}

@article{gebru2021datasheets,
  author        = {Gebru, Timnit and
               Morgenstern, Jamie and
               Vecchione, Briana and
               Vaughan, Jennifer Wortman and
               Wallach, Hanna and
               Daum{\'e}, III, Hal and
               Crawford, Kate},
  title         = {Datasheets for Datasets},
  journal       = {Communications of the {ACM}},
  volume        = {64},
  number        = {12},
  pages         = {86--92},
  year          = {2021},
  doi           = {10.1145/3458723},
  eprint        = {1803.09010},
  archivePrefix = {arXiv}
}

@inproceedings{akhtar2024croissant,
  author        = {Akhtar, Mubashara and
               Benjelloun, Omar and
               Conforti, Costanza and
               Foschini, Luca and
               Gijsbers, Pieter and
               Giner-Miguelez, Joan and
               Goswami, Sujata and
               Jain, Nitisha and
               Karamousadakis, Michalis and
               Krishna, Satyapriya and
               Kuchnik, Michael and
               Lesage, Sylvain and
               Lhoest, Quentin and
               Marcenac, Pierre and
               Maskey, Manil and
               Mattson, Peter and
               Oala, Luis and
               Oderinwale, Hamidah and
               Ruyssen, Pierre and
               Santos, Tim and
               Shinde, Rajat and
               Simperl, Elena and
               Suresh, Arjun and
               Thomas, Goeffry and
               Tykhonov, Slava and
               Vanschoren, Joaquin and
               Varma, Susheel and
               van der Velde, Jos and
               Vogler, Steffen and
               Wu, Carole-Jean and
               Zhang, Luyao},
  title         = {Croissant: A Metadata Format for {ML}-Ready Datasets},
  booktitle     = {Advances in Neural Information Processing Systems 37 (NeurIPS 2024), Datasets and Benchmarks Track},
  pages         = {82133--82148},
  year          = {2024},
  doi           = {10.52202/079017-2610},
  eprint        = {2403.19546},
  archivePrefix = {arXiv}
}

@misc{ritchie2026kimi,
  author       = {Ritchie, Logan and Mehta, Sushant and Panavas, Liudas and Chen, Edwin},
  title        = {Post-Training {Kimi-K2.7} on {GDP.pdf}: Benchmark Results and Transfer to Tool-Using Professional Tasks on {GDPval}},
  howpublished = {Surge AI technical report},
  month        = sep,
  year         = {2026}
}

@article{yu2025dapo,
  author  = {Yu, Qiying and Zhang, Zheng and Zhu, Ruofei and Yuan, Yufeng and Zuo, Xiaochen and Yue, Yu and Fan, Tiantian and Liu, Gaohong and Liu, Lingjun and Liu, Xin and others},
  title   = {{DAPO}: An Open-Source {LLM} Reinforcement Learning System at Scale},
  journal = {arXiv preprint arXiv:2503.14476},
  year    = {2025}
}

\clearpage
\appendix

\section{Solver Verification}
\label{app:verification}

\begin{figure}[H]
  \centering
  \includegraphics[width=\linewidth,height=0.3\textheight,keepaspectratio]{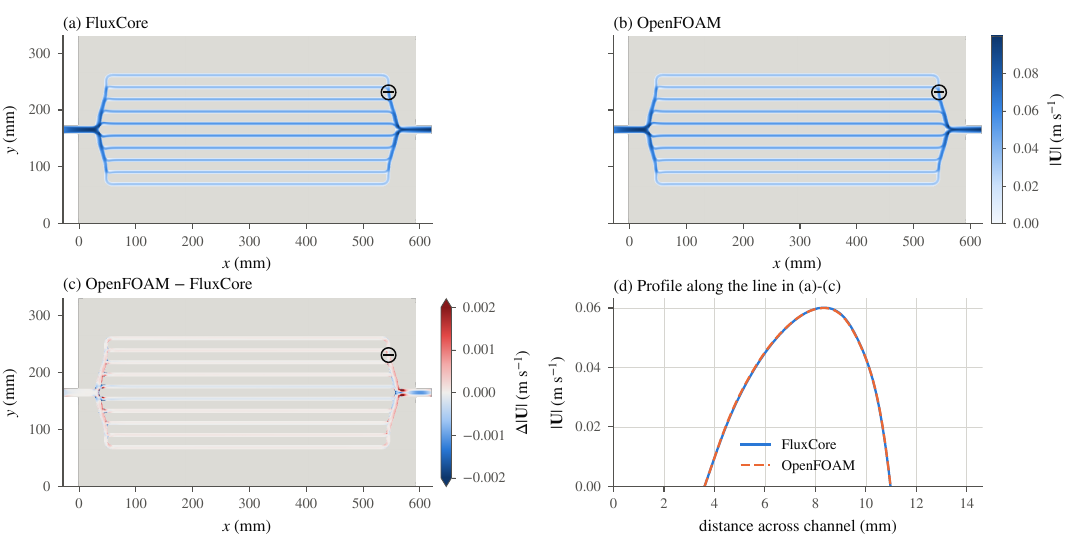}
  \caption{Coolant velocity magnitude on the mid-channel plane $z = 8.19$\,mm for verification case 3 (laminar, $\mathrm{Re} = 35$). (a) FluxCore; (b) OpenFOAM v2512; (c) OpenFOAM minus FluxCore, cell by cell on the shared mesh. The cold plate is grey. (d) $|\mathbf{U}|$ across the 7.31\,mm channel marked in (a)--(c), drawn as a smooth curve through the 10 cell-centre values with zero speed at the walls. Peak $|\mathbf{U}|$ on the plane is 0.100\,m/s for both solvers. Over all coolant cells the volume-weighted RMS difference is 0.0002\,m/s, 0.2\% of the peak speed.}
  \label{fig:case00203_coolant_velocity}
\end{figure}

\begin{figure}[H]
  \centering
  \includegraphics[width=\linewidth,height=0.3\textheight,keepaspectratio]{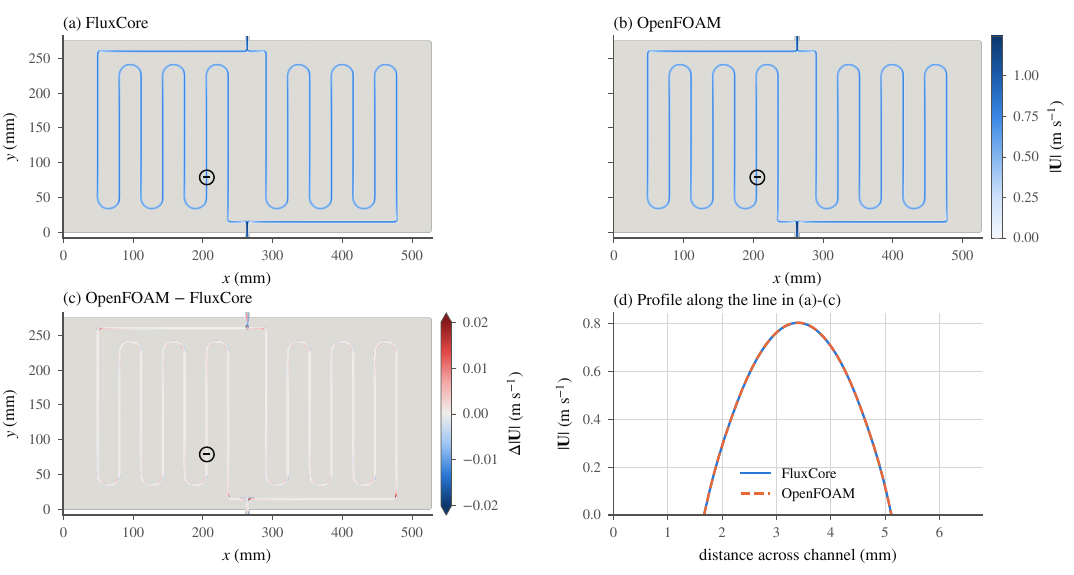}
  \caption{Coolant velocity magnitude on the mid-channel plane $z = 3.75$\,mm for verification case 11
  (laminar, $\mathrm{Re} = 292$). Panels as in Figure~\ref{fig:case00203_coolant_velocity}; (d) is
  across the 3.40\,mm channel, through 11 cell-centre values. Peak $|\mathbf{U}|$ on the plane is
  1.245\,m/s (FluxCore) and 1.246\,m/s (OpenFOAM). Over all coolant cells the volume-weighted RMS
  difference is 0.0015\,m/s, 0.1\% of the peak speed.}
  \label{fig:case06268_coolant_velocity}
\end{figure}

\begin{figure}[H]
  \centering
  \includegraphics[width=\linewidth,height=0.3\textheight,keepaspectratio]{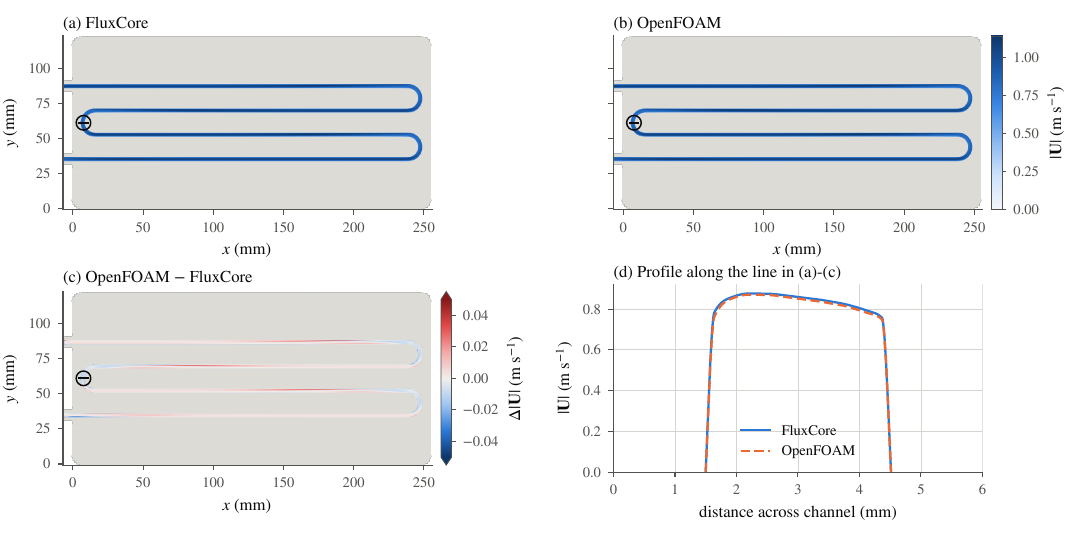}
  \caption{Coolant velocity magnitude on the mid-channel plane $z = 5.42$\,mm for verification case 20
  (SST $k$--$\omega$, $\mathrm{Re} = 2{,}331$). Panels as in
  Figure~\ref{fig:case00203_coolant_velocity}; (d) is across the 3.00\,mm channel, through 10
  cell-centre values. Peak $|\mathbf{U}|$ on the plane is 1.128\,m/s (FluxCore) and 1.140\,m/s
  (OpenFOAM). Over all coolant cells the volume-weighted RMS difference is 0.0073\,m/s, 0.6\% of the
  peak speed.}
  \label{fig:case08200_coolant_velocity}
\end{figure}

\begin{figure}[H]
  \centering
  \includegraphics[width=\linewidth,height=0.3\textheight,keepaspectratio]{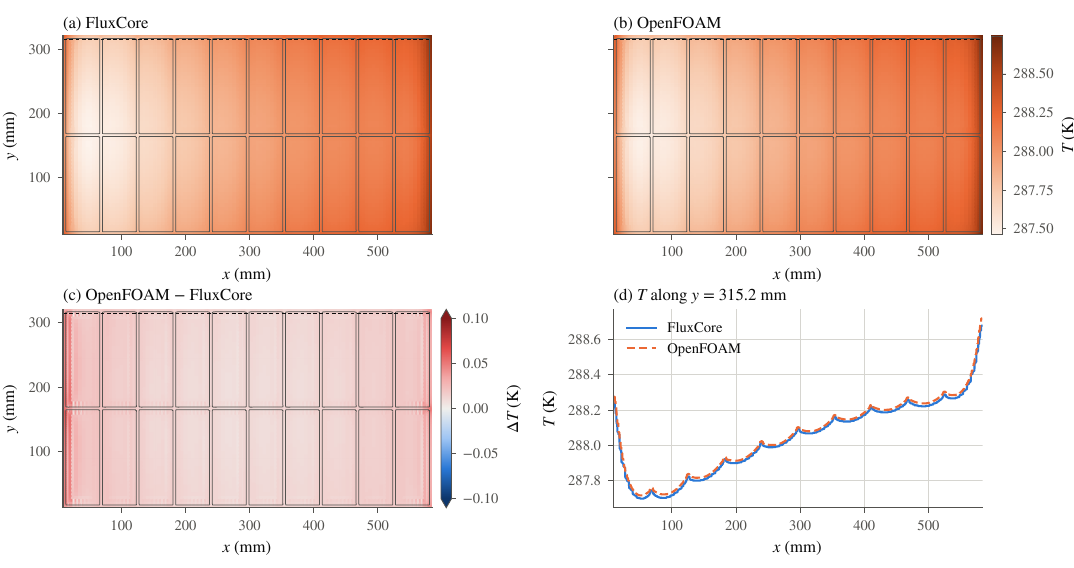}
  \caption{Temperature on the horizontal plane $z = 77.7$\,mm, at mid-height of the battery cells, for verification case 3 (laminar, $\mathrm{Re} = 35$). Cells and potting are shown, with cell outlines in grey. (a) FluxCore; (b) OpenFOAM; (c) OpenFOAM minus FluxCore. (d) Temperature along the dashed line $y = 315.2$\,mm, through the hottest point on the plane. Plane maximum: 288.61\,K (FluxCore), 288.65\,K (OpenFOAM). Over all battery-cell mesh cells, the volume-weighted mean difference is $+0.014$\,K and the RMS difference 0.016\,K.}
  \label{fig:case00203_battery_plane_temperature}
\end{figure}

\begin{figure}[H]
  \centering
  \includegraphics[width=\linewidth,height=0.3\textheight,keepaspectratio]{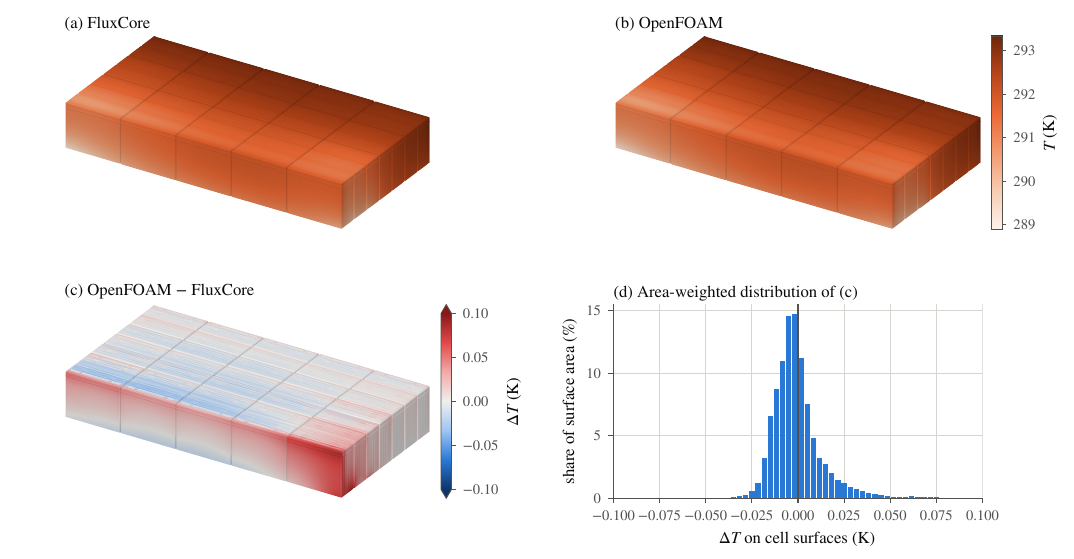}
  \caption{Temperature on the outer surfaces of the battery cells (621,777 mesh faces) for
  verification case 8 (laminar, $\mathrm{Re} = 103$), viewed from azimuth $-58^\circ$ and elevation
  $28^\circ$. (a) FluxCore; (b) OpenFOAM; (c) OpenFOAM minus FluxCore. (d) Area-weighted distribution
  of (c): mean $-0.001$\,K, RMS 0.014\,K, 5th to 95th percentile $-0.017$ to $+0.025$\,K.}
  \label{fig:case08454_cell_surface_temperature}
\end{figure}

This appendix extends Section~\ref{sec:answer-validity}. OpenFOAM v2512 was run on the FluxCore mesh
with the same boundary conditions, so the two fields can be differenced cell by cell. Statistics use
every verification case; pictures show released designs only.
Figures~\ref{fig:case00203_coolant_velocity}--\ref{fig:case08200_coolant_velocity} show coolant
velocity in two laminar cases and one turbulent case, and
Figures~\ref{fig:case00203_battery_plane_temperature} and~\ref{fig:case08454_cell_surface_temperature}
show cell temperature.

\section{Full Tasks}
\label{app:tasks}

\paragraph{Messages.} Every model gets the same system prompt: \emph{``You are an experienced thermal
engineer. Answer from engineering judgement and first-principles estimates; you cannot run
simulations.''} A single-turn task appends to its prompt: \emph{``Finish your reply with a single JSON
object with exactly these fields:} \texttt{<output schema>}\emph{. Write every interval as [low, high]
numbers and every confidence as a number between 0 and 1.''} An \texttt{evidence\_signoff} task
instead appends a turn protocol: one JSON tool call per reply,
\texttt{\{"tool": <name>, "args": \{...\}\}}, with the tool list and the budget, ending with
\texttt{submit}.

\paragraph{Tasks shown.} Figure~\ref{fig:task-campaign} is a campaign task from the released
\texttt{train} split and Figure~\ref{fig:task-evidence} an evidence task, both as stored with only
typesetting changed. Evidence tasks are multi-step, so none is in the released training files; the one
shown is from the training pool, on a \texttt{train} design. The envelope task of
Figure~\ref{fig:example} is released in full under its ID. Every task was built by generator 0.15.0;
each record also carries the canary string of Section~\ref{sec:leakage}.

\begin{figure}[p]
  \centering
  \fbox{\begin{minipage}{0.97\linewidth}\scriptsize
You are reviewing a test campaign on a battery cold plate. The five planned test points are listed below. Six result sheets came back from the rig, but their labels were lost and some entries are illegible. Five belong to these test points. One was misfiled from a test on a different plate. You may not run a simulation.\\[3pt]
\textbf{Design.}\\
Pack: 35 $\times$ EVE LF280K-class prismatic cell (280 Ah, LFP/graphite, 174 $\times$ 71.7 $\times$ 207 mm) in a rectangular grid of 7 rows $\times$ 5 columns, 1.2 mm cell spacing; a 1.0 mm thermal-interface pad under each cell; solid potting between cells up to 90\% of cell height (terminals and vents exposed).\\
Cold plate: Aluminium centre-fed parallel-channel plate, 900 $\times$ 536 $\times$ 14.5 mm; 16 parallel round channels fed by a bifurcating centre manifold; channel diameter 2.48 mm, flow along the plate length (x); ports 6.79 mm on the left (inlet) and right (outlet) faces; minimum wall 2.6 mm.\\[3pt]
\textbf{Planned test points.}\\
1: Coolant 30 wt\% ethylene glycol / water at 13.8\,\textdegree C inlet, 2.59 L/min (density 1040 kg/m$^3$, cp 3700 J/kgK, viscosity 2.63 mPa$\cdot$s). Ambient 17.8\,\textdegree C, external heat-transfer coefficient 8.2 W/m$^2$K on exposed surfaces. Load: discharge at 0.418C, SOC 89\%, state of health 89\%; total heat generation 445 W (397 W in cells, 48.3 W in terminals/interconnects, applied uniformly).\\
2: Coolant 50 wt\% ethylene glycol / water at 14.8\,\textdegree C inlet, 2.82 L/min (density 1068 kg/m$^3$, cp 3284 J/kgK, viscosity 4.41 mPa$\cdot$s). Ambient 23.7\,\textdegree C, external heat-transfer coefficient 5.8 W/m$^2$K on exposed surfaces. Load: discharge at 0.353C, SOC 39\%, state of health 75\%; total heat generation 726 W (716 W in cells, 9.67 W in terminals/interconnects, applied uniformly).\\
3: Coolant water at 12.5\,\textdegree C inlet, 0.500 L/min (density 999.4 kg/m$^3$, cp 4192 J/kgK, viscosity 1.22 mPa$\cdot$s). Ambient 31.7\,\textdegree C, external heat-transfer coefficient 11 W/m$^2$K on exposed surfaces. Load: charge at 0.251C, SOC 84\%, state of health 83\%; total heat generation 76.2 W (69.3 W in cells, 6.91 W in terminals/interconnects, applied uniformly).\\
4: Coolant water at 23.0\,\textdegree C inlet, 1.57 L/min (density 997.5 kg/m$^3$, cp 4182 J/kgK, viscosity 0.933 mPa$\cdot$s). Ambient 39.4\,\textdegree C, external heat-transfer coefficient 9.1 W/m$^2$K on exposed surfaces. Load: charge at 0.630C, SOC 60\%, state of health 87\%; total heat generation 375 W (351 W in cells, 23.9 W in terminals/interconnects, applied uniformly).\\
5: Coolant 50 wt\% ethylene glycol / water at 13.3\,\textdegree C inlet, 1.83 L/min (density 1068 kg/m$^3$, cp 3276 J/kgK, viscosity 4.65 mPa$\cdot$s). Ambient 31.1\,\textdegree C, external heat-transfer coefficient 6.8 W/m$^2$K on exposed surfaces. Load: charge at 0.526C, SOC 25\%, state of health 72\%; total heat generation 394 W (361 W in cells, 33.0 W in terminals/interconnects, applied uniformly).\\[3pt]
\textbf{Unlabelled result sheets (rises are above that run's coolant inlet).}\\
A peak cell rise 16.8 K; cell spread 14.7 K; coolant rise illegible; pressure drop 0.751 kPa\\
B peak cell rise 13.1 K; cell spread 12.0 K; coolant rise 2.11 K; pressure drop 10.7 kPa\\
C peak cell rise 17.0 K; cell spread 14.6 K; coolant rise 6.10 K; pressure drop illegible\\
D peak cell rise 19.7 K; cell spread 17.5 K; coolant rise illegible; pressure drop 10.6 kPa\\
E peak cell rise 21.1 K; cell spread 18.8 K; coolant rise 3.72 K; pressure drop 17.3 kPa\\
F peak cell rise 17.9 K; cell spread 16.4 K; coolant rise 4.48 K; pressure drop illegible\\[3pt]
\textbf{Sign-off: peak cell temperature must stay at or below the limit at each duty.}\\
Duty X (limit 33.0\,\textdegree C): Coolant 30 wt\% ethylene glycol / water at 23.6\,\textdegree C inlet, 0.500 L/min (density 1037 kg/m$^3$, cp 3729 J/kgK, viscosity 1.95 mPa$\cdot$s). Ambient 38.4\,\textdegree C, external heat-transfer coefficient 3.3 W/m$^2$K on exposed surfaces. Load: charge at 0.214C, SOC 44\%, state of health 98\%; total heat generation 26.2 W (25.0 W in cells, 1.19 W in terminals/interconnects, applied uniformly).\\
Duty Y (limit 51.5\,\textdegree C): Coolant water at 10.7\,\textdegree C inlet, 5.58 L/min (density 999.6 kg/m$^3$, cp 4194 J/kgK, viscosity 1.28 mPa$\cdot$s). Ambient 10.0\,\textdegree C, external heat-transfer coefficient 11 W/m$^2$K on exposed surfaces. Load: discharge at 0.859C, SOC 66\%, state of health 81\%; total heat generation 2201 W (2155 W in cells, 46.0 W in terminals/interconnects, applied uniformly).\\[3pt]
State which sheet belongs to each test point, which sheet is misfiled, and whether each duty passes.\\[3pt]
Scoring: The score is 1 only if every link, the misfiled sheet and both sign-offs are correct, and 0 otherwise. Your confidence does not change the score; it is only used to measure calibration.\par\vspace{3pt}
  \noindent\rule{\linewidth}{0.4pt}\par\vspace{1pt}
  \textbf{Output schema.} \texttt{confidence}; \texttt{links}: test point 1--5 $\to$ sheet;
  \texttt{misfiled}: sheet; \texttt{signoff}: X, Y $\to$ pass or fail.\\
  \textbf{Answer.} Links 1\,B, 2\,E, 3\,A, 4\,F, 5\,D; misfiled C; X fail (peak 34.2\,\textdegree C   against 33.0), Y pass (47.5 against 51.5).
  \end{minipage}}
  \caption{Campaign task \texttt{ejb-training-0.1-train-campaign-03763-c7768b}.}
  \label{fig:task-campaign}
\end{figure}

\begin{figure}[p]
  \centering
  \fbox{\begin{minipage}{0.97\linewidth}\scriptsize
You are signing off a battery cold plate at three new duties. You may not run a simulation. Stored results exist for the operating points below, but they are hidden. You can reveal at most 3 of them with get\_run\_result, so choose the runs that will tell you the most.\\[3pt]
\textbf{Design.}\\
Pack: 36 $\times$ 21700 cylindrical cell (Molicel P42A class) (4.20 Ah, INR layered oxide, 21.7 mm diameter $\times$ 70.2 mm tall) in a rectangular grid of 6 rows $\times$ 6 columns, 4.0 mm cell spacing; a 1.2 mm thermal-interface pad under each cell; open air between cells (natural convection and radiation to ambient).\\
Cold plate: Aluminium centre-fed parallel-channel plate, 179 $\times$ 178 $\times$ 15.0 mm; 4 parallel round channels fed by a bifurcating centre manifold; channel diameter 3.55 mm, flow along the plate length (x); ports 4.70 mm on the left (inlet) and right (outlet) faces; minimum wall 2.5 mm.\\[3pt]
\textbf{Stored runs (inputs only).}\\
R1: Coolant 50 wt\% ethylene glycol / water at 13.6\,\textdegree C inlet, 0.100 L/min (density 1068 kg/m$^3$, cp 3278 J/kgK, viscosity 4.59 mPa$\cdot$s). Ambient 11.9\,\textdegree C, external heat-transfer coefficient 11 W/m$^2$K on exposed surfaces. Load: charge at 0.640C, SOC 93\%, state of health 71\%; total heat generation 16.4 W (15.3 W in cells, 1.09 W in terminals/interconnects, applied uniformly).\\
R2: Coolant 50 wt\% ethylene glycol / water at 18.1\,\textdegree C inlet, 0.100 L/min (density 1066 kg/m$^3$, cp 3302 J/kgK, viscosity 3.94 mPa$\cdot$s). Ambient 36.0\,\textdegree C, external heat-transfer coefficient 11 W/m$^2$K on exposed surfaces. Load: discharge at 0.214C, SOC 56\%, state of health 76\%; total heat generation 1.94 W (1.86 W in cells, 0.0842 W in terminals/interconnects, applied uniformly).\\
R3: Coolant water at 17.9\,\textdegree C inlet, 0.100 L/min (density 998.6 kg/m$^3$, cp 4186 J/kgK, viscosity 1.06 mPa$\cdot$s). Ambient 37.7\,\textdegree C, external heat-transfer coefficient 3.9 W/m$^2$K on exposed surfaces. Load: discharge at 0.315C, SOC 25\%, state of health 98\%; total heat generation 8.68 W (8.66 W in cells, 0.0286 W in terminals/interconnects, applied uniformly).\\
R4: Coolant 30 wt\% ethylene glycol / water at 14.9\,\textdegree C inlet, 0.100 L/min (density 1040 kg/m$^3$, cp 3703 J/kgK, viscosity 2.53 mPa$\cdot$s). Ambient 14.0\,\textdegree C, external heat-transfer coefficient 4.4 W/m$^2$K on exposed surfaces. Load: charge at 0.521C, SOC 41\%, state of health 94\%; total heat generation 4.13 W (3.86 W in cells, 0.270 W in terminals/interconnects, applied uniformly).\\
R5: Coolant water at 20.8\,\textdegree C inlet, 0.673 L/min (density 998.0 kg/m$^3$, cp 4183 J/kgK, viscosity 0.981 mPa$\cdot$s). Ambient 26.4\,\textdegree C, external heat-transfer coefficient 7.5 W/m$^2$K on exposed surfaces. Load: discharge at 2.81C, SOC 47\%, state of health 79\%; total heat generation 177 W (174 W in cells, 3.37 W in terminals/interconnects, applied uniformly).\\[3pt]
\textbf{Duties.}\\
Duty X (peak cell limit 60.5\,\textdegree C, pressure drop limit 11.4 kPa): Coolant 30 wt\% ethylene glycol / water at 28.3\,\textdegree C inlet, 2.50 L/min (density 1034 kg/m$^3$, cp 3742 J/kgK, viscosity 1.71 mPa$\cdot$s). Ambient 41.1\,\textdegree C, external heat-transfer coefficient 8.6 W/m$^2$K on exposed surfaces. Load: discharge at 4.27C, SOC 78\%, state of health 86\%; total heat generation 188 W (180 W in cells, 7.57 W in terminals/interconnects, applied uniformly).\\
Duty Y (peak cell limit 15.5\,\textdegree C, pressure drop limit 0.0745 kPa): Coolant water at 9.3\,\textdegree C inlet, 0.100 L/min (density 999.8 kg/m$^3$, cp 4196 J/kgK, viscosity 1.33 mPa$\cdot$s). Ambient 20.1\,\textdegree C, external heat-transfer coefficient 5.3 W/m$^2$K on exposed surfaces. Load: charge at 0.400C, SOC 19\%, state of health 83\%; total heat generation 3.31 W (3.09 W in cells, 0.224 W in terminals/interconnects, applied uniformly).\\
Duty Z (peak cell limit 17.0\,\textdegree C, pressure drop limit 10.4 kPa): Coolant 30 wt\% ethylene glycol / water at 10.5\,\textdegree C inlet, 2.00 L/min (density 1042 kg/m$^3$, cp 3690 J/kgK, viscosity 2.93 mPa$\cdot$s). Ambient 29.1\,\textdegree C, external heat-transfer coefficient 8.8 W/m$^2$K on exposed surfaces. Load: discharge at 0.273C, SOC 89\%, state of health 95\%; total heat generation 1.18 W (1.06 W in cells, 0.122 W in terminals/interconnects, applied uniformly).\\[3pt]
For each duty, say whether peak cell temperature stays at or below its limit and whether pressure drop stays at or below its limit.\\[3pt]
Scoring: The score is 1 only if every temperature and pressure verdict is correct, and 0 otherwise; which runs you reveal is not scored. Your confidence does not change the score; ...it is only used to measure calibration.\par\vspace{3pt}
  \noindent\rule{\linewidth}{0.4pt}\par\vspace{1pt}
  \textbf{Reference episode.} Each line is one turn; tool results are shown as returned.\\
  \texttt{\{"tool": "get\_run\_result", "args": \{"run\_id": "R1"\}\}}\\
  \quad$\to$ \texttt{\{"run\_id": "R1", "cell\_spread\_K": 3.25, "coolant\_rise\_K": 1.596,
    "dP\_kPa": 0.1567, "dT\_max\_K": 5.009, "max\_cell\_C": 18.65\}}\\
  \texttt{\{"tool": "get\_run\_result", "args": \{"run\_id": "R3"\}\}}\\
  \quad$\to$ \texttt{\{"run\_id": "R3", "cell\_spread\_K": 7.414, "coolant\_rise\_K": 3.792,
    "dP\_kPa": 0.05547, "dT\_max\_K": 11.096, "max\_cell\_C": 29.01\}}\\
  \texttt{\{"tool": "get\_run\_result", "args": \{"run\_id": "R5"\}\}}\\
  \quad$\to$ \texttt{\{"run\_id": "R5", "cell\_spread\_K": 26.129, "coolant\_rise\_K": 3.242,
    "dP\_kPa": 0.9692, "dT\_max\_K": 33.883, "max\_cell\_C": 54.73\}}\\
  \texttt{\{"tool": "submit", "args": \{"temperature": \{"X": "fail", "Y": "pass", "Z": "fail"\},
    "pressure": \{"X": "pass", "Y": "pass", "Z": "pass"\}, "confidence": \ldots\}\}}\\[2pt]
  \textbf{Answer.} Temperature X fail (peak 64.7\,\textdegree C against 60.5), Y pass (14.9 against
  15.5), Z fail (17.9 against 17.0); pressure drop passes at all three (10.1, 0.0639 and
  8.80\,kPa).
  \end{minipage}}
  \caption{Evidence task \texttt{ejb-training-0.1-train-evsign-01153-1b5d78}, from the training pool.
  Revealing the three runs most like the duties (R2, R3, R4) leads the reference fit to a wrong
  verdict; the reference set R1, R3, R5 spans the inputs and gets every verdict right, as do four of
  the ten possible choices of three runs.}
  \label{fig:task-evidence}
\end{figure}

\paragraph{Training-only format.} \texttt{duty\_signoff} (Appendix~\ref{app:training}) has the
evidence layout with every shown run's results visible, one or two duties, and no tools. It is
single-turn and scored like \texttt{evidence\_signoff}.

\section{Training Details}
\label{app:training}

\begin{table}[ht]
  \centering
  \caption{The released training set (\texttt{training-0.1}), tasks by format. It is a subset of a
  pool of 4,459 training and 446 validation tasks: multi-step tasks and tasks a quick method solves
  (for envelopes, ranking by one input or a quick estimate; for sign-offs, the closest run alone) are
  left out, and each design keeps at most three campaign or envelope tasks and one sign-off task per
  format (seed 20261109).}
  \label{tab:training-set}
  \small
  \begin{tabular}{llrr}
    \toprule
    Family & Format & Train & Val \\
    \midrule
    \texttt{campaign\_reconstruction} & 5 points          & 300 & 30 \\
                                      & 4 points          & 317 & 32 \\
    \texttt{operating\_envelope}      & 4 duties          & 98  & 10 \\
                                      & 3 duties          & 53  & 7 \\
    \texttt{duty\_signoff}            & 1 duty, 3 runs    & 269 & 30 \\
                                      & 1 duty, 4 runs    & 111 & 11 \\
                                      & 2 duties, 3 runs  & 133 & 13 \\
    \midrule
    All     & & 1,281 & 133 \\
    Designs & & 289   & 31 \\
    \bottomrule
  \end{tabular}
\end{table}

\paragraph{Reward.} The reward is the fraction of criteria met, checked by the benchmark scorer: each link, the misfiled sheet and each verdict in a campaign task; each temperature and pressure verdict in a sign-off task; the ranking (exact) and each regime in an envelope task. A full pass earns no extra reward. A reply with no answer the scorer can read earns $-1$, and a reply within 4,096 tokens of the length limit loses up to 1 more, falling linearly to the limit~\citep{yu2025dapo}. Half of the training prompts also ask for single-number estimates behind the verdicts (the peak cell rise of each duty, and the pressure drop or ambient-gain ratio where the family uses one). Each estimate scores 1 within its task tolerance and falls linearly to 0 at five tolerances, and on those prompts the reward is $0.7 \times$ the criteria reward $+\, 0.3 \times$ the estimate score. Evaluation always uses the benchmark prompt. Only the text after the last \texttt{</think>} is scored.

\paragraph{Task selection.} Every training task is eligible. Prompts are drawn in a seeded random order, fixed before training, so that each family supplies a third of them. This repeats the 151 envelope tasks (2.8 times each on average), and a full pass over the resulting 1,851 prompts would take 232 steps. The 160-step budget covers 953 distinct tasks.

\begin{table}[ht]
  \centering
  \caption{Training settings.}
  \label{tab:training-settings}
  \small
  \begin{tabular}{@{}lp{0.74\linewidth}@{}}
    \toprule
    Setting & Value \\
    \midrule
    Model & Qwen3.5-9B, thinking on \\
    Algorithm & TRL~\citep{vonwerra2020trl}: advantages normalised within each group; sequence-level
      importance ratio clipped at $3\times10^{-4}$ / $4\times10^{-4}$ (GSPO~\citep{zheng2025gspo});
      loss averaged over every reply token in the batch (DAPO); KL coefficient 0.02 towards the base
      model; one update per batch \\
    Group and batch & 8 replies per task, 8 tasks per step, 160 steps \\
    Optimiser & Adam (0.9, 0.98), learning rate $5\times10^{-5}$ constant, weight decay 0.1,
      gradient clip 0.1 \\
    Adapter & LoRA~\citep{hu2022lora}, rank 32, $\alpha$ 32, dropout 0, all linear layers outside the
      vision encoder (86.6M trainable parameters, 0.91\%) \\
    Sampling & temperature 0.6, top-$p$ 0.95, top-$k$ 20; replies up to 32,768 tokens \\
    Generation & vLLM~\citep{kwon2023vllm} on the training GPU, bf16 \\
    Hardware & one B200, about 9 minutes per step; seed 0 \\
    Checkpoint & step 160, fixed before any test evaluation \\
    Evaluation & base and trained models four times each; temperature 0.6, top-$p$ 0.95, top-$k$ 20,
      65,536 tokens \\
    \bottomrule
  \end{tabular}
\end{table}

The settings start from the second stage of a published recipe for rubric-rewarded
RL~\citep{ritchie2026kimi}, with 8 tasks per update instead of 64 to fit one GPU. Four changes came
from pilot runs on \texttt{train} and \texttt{validation}:
\begin{itemize}
  \item The recipe's clip range of 0.20/0.28 never bound for a sequence-level ratio, so we use GSPO's
  range.
  \item At its learning rate of $10^{-5}$ the adapter barely moved, so we use $5\times10^{-5}$.
  \item Averaging each reply's loss separately shortened replies, so we average over all reply tokens.
  \item A KL term holds the gain on \texttt{validation}, which otherwise drifted away after about
  12 steps.
\end{itemize}
Training also samples at the evaluation settings, because replies sampled at temperature 0.7 stayed
within the limit in training but often ran past it in evaluation.

\section{Datasheet}
\label{app:datasheet}

Answers follow the questions of \citet{gebru2021datasheets}, shortened.

\paragraph{Motivation.} EJBench measures whether a model can predict a design's behaviour at new
conditions from a few earlier results and decide on it. It was created and funded by Gradient
Dynamics.

\paragraph{Composition.} Each instance is one task: a prompt, an output schema, the answer, a scoring rule, the stored inputs, tags and provenance, plus, for multi-step tasks, the hidden runs served by the tool. The released splits are \texttt{train} (1,281 tasks), \texttt{validation} (133) and \texttt{test} (\nTestTasks{}, of which \nScoredTasks{} are scored; Tables~\ref{tab:splits} and~\ref{tab:training-set}). Every task comes from one design and its simulations, apart from the misfiled sheet of a campaign task. The designs are generated by a design of experiments; no instance relates to a person, and the data hold no personal information. Cell types are named by class (for example ``EVE LF105-class''). A private split is withheld as a contamination check (Section~\ref{sec:data}).

\paragraph{Uses.} \texttt{test} is for evaluation only and must not be trained on. \texttt{train} and \texttt{validation} may be used for development and training. EJBench does not certify designs: a score is not evidence that a model can sign off real hardware.

\end{document}